\documentclass[preprint,notoc]{JHEP3}
\usepackage{epsfig}

\def\to{\rightarrow}

\def\bi{\begin{itemize}}
 \def\ei{\end{itemize}}
\def\te{\tilde e}
\def\c1p{C1^\prime}
\def\ta{\tilde a}
\def\tG{\widetilde G}

\def\tu{\tilde u}

\def\ta{\tilde a}

\def\tb{\tilde b}

\def\tst{\tilde t}

\def\tw{\widetilde W}
\def\tz{\widetilde Z}
\def\mgut{M_{\rm GUT}}
\def\alt{\lesssim}
\def\agt{\gtrsim}
\def\be{\begin{equation}}  
\def\ee{\end{equation}}  
\def\bea{\begin{eqnarray}}  
\def\eea{\end{eqnarray}}  

\newcommand\njp[3]{{\it New\ J.\ Phys.\ }{\bf #1} (#2) #3}
\newcommand\ijmpe[3]{{\it Int.\ J.\ Mod. Phys.\ }{\bf E#1} (#2) #3}
\newcommand\annp[3]{{\it Annals\ Phys.\ }{\bf #1} (#2) #3}
\newcommand\sjp[3]{{\it Sov.\ J.\ Nucl.\ }{\bf #1} (#2) #3}

\def\Isajet{{\sc Isajet}}

\title{Reconciling thermal leptogenesis with  
the gravitino problem in SUSY models
with mixed axion/axino dark matter}

\author{Howard Baer$^{a}$, Sabine Kraml$^b$, Andre Lessa$^{a}$ and 
Sezen Sekmen$^c$\\
$^a$Dept.\ of Physics and Astronomy, University of Oklahoma, Norman, OK 73019, USA\\
$^b$Laboratoire de Physique Subatomique et de Cosmologie, UJF Grenoble 1, 
CNRS/IN2P3, INPG, 53 Avenue des Martyrs, F-38026 Grenoble, France\\
$^c$Dept.\ of Physics, Florida State University, Tallahassee, FL 32306, USA\\
E-mail: \email{baer@nhn.ou.edu}, \email{sabine.kraml@lpsc.in2p3.fr}, 
\email{lessa.a.p@gmail.com}, \email{sezen.sekmen@cern.ch}}

\abstract{
Successful implementation of thermal leptogenesis requires re-heat
temperatures $T_R\agt 2\times 10^9$ GeV, in apparent conflict with 
SUSY models with TeV-scale gravitinos, 
which require much lower $T_R$ in order to avoid 
Big Bang Nucleosynthesis (BBN) constraints. 
We show that mixed axion/axino dark matter can reconcile 
thermal leptogenesis with the gravitino problem in models with 
$m_{\tG}\agt 30$ TeV, a rather high Peccei-Quinn breaking scale
and an initial mis-alignment angle $\theta_i < 1$.
We calculate axion and axino dark matter production from four sources, 
and impose BBN constraints on long-lived gravitinos and neutralinos.
Moreover, we discuss several SUSY models which naturally have 
gravitino masses of the order of tens of TeV.
We find a reconciliation difficult in Yukawa-unified SUSY 
and in AMSB with a wino-like lightest neutralino. 
However, $T_R\sim 10^{10}-10^{12}$~GeV can easily be achieved in 
effective SUSY and in models based on mixed moduli-anomaly mediation. 
Consequences of this scenario include: 
1.~an LHC SUSY discovery should be consistent
with SUSY models with a large gravitino mass,
2.~an apparent neutralino relic abundance $\Omega_{\tz_1}h^2\alt 1$,
3.~no WIMP direct or indirect detection signals should be found, and 
4.~the axion mass should be less than $\sim 10^{-6}$~eV, somewhat below 
the conventional range which is explored by microwave cavity axion 
detection experiments.
}

\begin{document}

\section{Introduction}
\label{sec:intro}

Recent measurements of neutrino oscillations~\cite{nu_review} are elegantly 
interpreted in terms of see-saw neutrino masses~\cite{seesaw}, wherein heavy 
right-handed neutrino (RHN) states $N_i$ ($i=1-3$ for three generations) 
are introduced, and the light neutrino masses are given approximately by 
$m_{\nu_i}\simeq m_{D_i}^2/M_{N_i}$, where $m_{Di}\sim f_{\nu_i} v$ with 
$f_{\nu_i}$ the neutrino Yukawa coupling and $v$ the Higgs field vacuum 
expectation value.
A value of $M_{N_i}\sim 10^{15}$ GeV yields $m_{\nu_\tau}\sim 0.03$~eV
in the GUT-inspired case where $f_{\nu_\tau}=f_t$ at $\mgut$.

One of the appealing consequences of such heavy RHN states is that 
the baryon number of the universe can be explained in terms of 
thermal leptogenesis~\cite{lepto_review}. 
In thermal leptogenesis, the right-hand neutrinos are present in 
thermal equilibrium at high temperatures in the early universe, and 
decay asymmetrically to leptons versus anti-leptons. The lepton asymmetry 
is converted to a baryon asymmetry via $B$- and $L$-violating but 
$B-L$ conserving sphaleron interactions. 
In a scheme with hierarchical right-hand neutrino masses, 
a re-heat temperature 
\be
   T_R\agt 2\times 10^9\ {\rm GeV}\ \ \ (\mbox{{\rm thermal leptogenesis}})
\ee
is required to reproduce the observed baryon asymmetry of the universe~\cite{TR}.

The existence of a heavy mass scale like $M_N$ or $\mgut$ brings about the 
infamous hierarchy problem of the Standard Model (SM). 
The gauge hierarchy problem is elegantly resolved by introducing 
supersymmetry (SUSY) into the theory.
SUSY is a novel spacetime symmetry which relates bosons and fermions, see {\it e.g.}~\cite{wss}.
In realistic models, SUSY is broken ``softly'' at the weak scale---implying 
that all SM particles must have superpartners with masses in the range up 
to ${\cal O}(1)$~TeV that ought to be accessible to colliders such as the CERN LHC.
SUSY not only stabilizes the hierachy between weak scale and other high scales such as
$\mgut$; indeed with weak-scale SUSY the gauge couplings, when evolved upwards 
from $Q=M_Z$ under renormalization group evolution (RGE), are found to unify at 
$Q\approx 2\times 10^{16}$~GeV, which is nicely consistent with the idea of a 
grand unified theory.
Sensible implementations of supersymmetry invoke SUSY as a local 
symmetry (supergravity or SUGRA), which requires in addition the existence of
a graviton/gravitino supermultiplet. In SUGRA models, supergravity is broken via
the superHiggs mechanism, leading to a massive gravitino $\tG$. The 
soft SUSY breaking mass terms are expected to be closely related
to the gravitino mass $m_{\tG}$, and so $m_{\tG}$ is also 
expected to be of the order of the weak scale. 

One of the impediments to successful SUGRA model building is known
as the gravitino problem~\cite{gravprob}. Gravitinos can be produced thermally
in the early universe, even though their Planck-suppressed couplings
preclude them from participating in thermal equilibrium. 
The gravitino decay rate is also suppressed by the Planck scale, 
leading to very long gravitino lifetimes of order $1-10^5$ sec. 
There are actually two parts to the gravitino problem. 
Part 1 is that for $m_{\tG}\sim 1$ TeV, the late time
gravitino decays can inject hadronic or electromagnetic energy into  the
cosmic plasma at a time scale during or after Big Bang Nucleosynthesis (BBN),
leading to destruction of the successful agreement between theory and
observation for the light element abundances. 
For re-heat temperatures $T_R\alt 10^5$ GeV, 
thermal gravitino production is suppressed enough to evade BBN limits~\cite{kl}.
However, these low of temperatures are in conflict with those needed for
thermal leptogenesis. 

If $5\ {\rm TeV}\lesssim m_{\tG}\lesssim 30$~TeV, then the gravitino 
lifetime drops below 1~sec, and the BBN constraints are much more
mild, allowing $T_R\alt 10^9$~GeV. This range of $T_R$ is consistent
with non-thermal leptogenesis~\cite{ntlepto}, wherein other sources of 
$N_i$, such as inflaton decay, contribute to $N_i$ production. 
For heavier yet gravitinos with $m_{\tG}\agt 30$ TeV, 
the value of $T_R$ can reach as high as $7\times 10^{9}$ GeV. 
In this case---part 2 of the gravitino problem---the upper bound on 
$T_R$ comes from overproduction of neutralino dark matter due to 
their combined thermal production and production via gravitino decay.
Models such as AMSB, with multi-TeV gravitino masses and very low thermal
neutralino abundances, thus naturally reconcile thermal leptogenesis
with the gravitino problem, but only for the narrow range
$2\times 10^9\ {\rm GeV}\alt T_R\alt 7\times 10^{9}$ GeV~\cite{amsb_soln}.

One way out is to invoke a gravitino as LSP, with a stau or
neutralino as the next-to-LSP 
(NLSP) which decays via a small $R$-parity violating interaction~\cite{covi} 
(in the R-parity conserving case, it is very difficult to reconcile 
gravitino DM with thermal leptogenesis~\cite{covi,covi2}). 
The gravitino, which may also decay via $R$-parity
violating interactions, has a lifetime of order the age of the universe; 
it can still function as dark matter, but its
occasional decays in the galactic halo could be the source of Pamela, ATIC  
and Fermi cosmic ray anomalies~\cite{cr_buch}.

Another way out involves mixed axion/axino DM, and this is the topic of this paper. 
Indeed, the strong $CP$ problem remains as one of the
central puzzles of QCD which evades explanation within the context of
the Standard Model. The crux of the problem is that an additional
$CP$ violating term in the QCD Lagrangian of the form\footnote{Here $G_{\mu\nu}^a$ 
is the gluon field strength tensor and $\tilde{G}^{a\mu\nu}$ its dual.} 
$\bar{\theta}g_s^2/32\pi^2 G^{\mu\nu}_A \widetilde{G}_{A\mu\nu}^{}$ 
ought to be present as a result of the t'Hooft resolution of the $U(1)_A$
problem via instantons and the $\theta$ vacuum of QCD~\cite{thooft}. 
The experimental limits on the neutron electric dipole moment
however constrain $|\bar{\theta}|<10^{-10}$~\cite{nedm}. 
Why this term should be so small is the essence of the strong $CP$ problem.

An extremely compelling solution proposed by Peccei and Quinn~\cite{pq} is
to hypothesize an additional global $U(1)_{\rm PQ}$ symmetry, which is broken at
some high mass scale $f_a$.
A consequence of the broken PQ symmetry is the existence of a 
pseudo-Goldstone boson field: the axion $a(x)$~\cite{ww}. 
The Lagrangian then also contains the terms
\be
{\cal L}\ni {1\over 2}\partial_\mu a\,\partial^\mu a +\frac{g^2}{32\pi^2}\frac{a(x)}{f_a/N}
G_{\mu\nu}^a \tilde{G}^{a\mu\nu} \,, \label{L_ax}
\ee
where $N$ is the model-dependent color anomaly factor, which is $1$ for KSVZ~\cite{ksvz}
or $6$ for DFSZ~\cite{dfsz} models.  
Since $a(x)$ is dynamical, the entire $CP$-violating term settles to its
minimum at zero, thus resolving the strong $CP$ problem. 
A consequence of this very elegant mechanism is that
a physical axion field should exist, with axion excitations of mass~\cite{bardeen}
\be
m_a\simeq 6\ {\rm eV}\ \frac{10^6\ {\rm GeV}}{f_a/N} \, .
\label{eq:axmass}
\ee
The axion field couples to gluon-gluon (obvious from Eq. (\ref{L_ax})) and also to photon-photon 
and fermion-fermion. All the couplings are suppressed by the PQ scale $f_a$. 
The value of $f_a$ is constrained to lie above $\sim 10^9$ GeV by stellar cooling 
arguments~\cite{dicus}, leading to a nearly invisible axion particle which may be
searched for via microwave cavity experiments~\cite{sikivie}. 
In addition, axions can be produced via various mechanisms in the early universe. 
Since their lifetime (they decay via $a\to\gamma\gamma$) turns out to 
be longer than the age of the universe, they can be a good candidate for the  
DM of the universe~\cite{axion}.

In the context of supersymmetry, the axion field is but one element of an 
axion supermultiplet which also contains an $R$-parity even spin-0 saxion field 
$s(x)$ and an $R$-parity odd spin-${1\over 2}$ axino field $\ta$\cite{kim}. 
The axino field $\ta$ may play a huge role in cosmology~\cite{rtw}: 
its mass may lie anywhere in the range of keV to TeV~\cite{amass},
and it may function as the LSP. As the LSP, in $R$-parity conserving 
models, it may constitute at least a portion of the DM of the 
universe~\cite{ckkr,axino}.
The saxion field may also play a role in cosmology, {\it e.g.} via dilution of 
relics by additional entropy production~\cite{st,hasenkamp}, although we will 
not consider this here.

In~\cite{ay}, Asaka and Yanagida proposed to reconcile thermal leptogenesis
with the gravitino problem by requiring an axino LSP with a gravitino NLSP.
In this paper, we present a different reconciliation of thermal leptogenesis
with the gravitino problem, {\it nota bene} for very heavy gravitinos. 
First, we require $m_{\tG}\agt 30$ TeV so as to avoid part~1 of the 
gravitino problem. Next, we invoke the presence of mixed axion/axino 
dark matter into our scenario, which arises as a result of the PQ solution 
to the strong CP problem in the supersymmetric context~\cite{nilles}.
We assume here that the lightest neutralino, $\tz_1$, is the NLSP, 
so that each neutralino ultimately decays to an axino, and the neutralino 
relic mass abundance is then suppressed by a factor of $m_{\ta}/m_{\tz_1}$,
thus avoiding part~2 of the gravitino problem, the overproduction of 
neutralino dark matter.

In Sec.~\ref{sec:Oh2} of this paper, we evaluate the relic abundance of 
mixed axion/axino DM due to four sources. 
The first is ordinary production of axion cold DM via the vacuum mis-alignment 
mechanism. The second is thermal production (TP) of axinos: 
here, we restrict $T_R$ to values below the axino decoupling temperature ($T_{dcp}$), 
so axinos are never in thermal equilibrium. Nonetheless, axinos can still be produced 
thermally via bremsstrahlung and decays of particles which are in thermal equilibrium: 
the final result depends linearly on the re-heat temperature $T_R$ after inflation. 
The third is non-thermal production (NTP) of axinos via thermal neutralino production 
and decay. The fourth, also NTP of axinos comes from thermal production of 
gravitinos, followed by their cascade decays to the axino LSP state;
this mechanism also depends linearly on $T_R$.

For low values of PQ breaking scale $f_a/N\sim 10^9-10^{11}$, 
the axino coupling to matter is large
enough that thermal production of axinos tends to dominate the mixed axion/axino abundance. 
An upper bound on $T_R$ can be extracted by requiring the axino DM abundance lie below 
the WMAP-measured value~\cite{wmap7}:
\be
\Omega_{\rm DM}h^2=0.1123\pm 0.0035\ \ \ {\rm at\ 68\%\ CL} .
\ee
As $f_a/N$ increases, the portion of axino DM decreases, while the axion CDM increases.
Simplistic estimates of the relic axion abundance assume an initial mis-alignment
angle $\theta_i\simeq 1$, leading to an upper bound on PQ breaking scale 
$f_a/N\alt 5\times 10^{11}$~GeV. By adopting a smaller value of $\theta_i$, much larger 
values of $f_a/N$ in the $10^{13}10^{14}$~GeV range become allowed. This in turn suppresses 
the axino composition of DM, unless very high values of $T_R\agt 10^{10}$~GeV are allowed. 
This is the crux of our reconcilation of thermal leptogenesis with the gravitino problem.

However, within this solution, another BBN bound emerges, since 
the lifetime of the lightest neutralino scales as $(f_a/N)^{-2}$.
There exist additional strict limits on late decaying particles with 
electromagnetic and hadronic energy injection into the thermal plasma during BBN. 
Therefore, these limits provide an upper bound on $f_a/N$ for SUSY models 
including the PQ mechanism. In Sec.~\ref{sec:z1}, we evaluate the neutralino
lifetime and hadronic branching fraction, so that BBN constraints can be applied to $\tz_1$ decay.

In Sec.~\ref{sec:TR}, we present five scenarios which are consistent with gravitinos 
of mass $\agt 30$ TeV. The first two cases come from  gravity mediated SUSY breaking: 
the Yukawa-unified (YU) models and effective SUSY (ESUSY) models: both of these require 
GUT scale scalar masses in the $10-30$~TeV regime, and so should be consistent with 
gravitinos in this range. As we will see, the rather large abundance of bino-like 
neutralinos from YU models typically makes them inconsistent with $T_R$ values in 
excess of $2\times 10^9$ GeV. 
ESUSY models can more easily accommodate a low production rate for neutralinos in 
the early universe, and do allow for $T_R>10^{10}$ GeV.
The third case, that of anomaly-mediated SUSY breaking (AMSB), requires gravitinos in 
the $30-100$ TeV range since soft SUSY breaking terms are loop suppressed. Most versions 
of AMSB include a nearly pure wino-like neutralino. Since only the bino component of 
$\tz_1$ couples to the axino, the neutral wino-like $\tz_1$ decay is suppressed, and just 
barely allows $T_R> 2\times 10^9$ GeV, before BBN constraints kick in.
The fourth and fifth benchmark points come from ``mirage unification'' (MU), or 
mixed moduli-anomaly mediation, which also allows $m_{\tG}\sim 30-100$ TeV. These models 
can easily include a bino-like $\tz_1$ with a relatively low relic abundance, avoiding 
the worst of BBN constraints. In these models, values of $T_R\sim 10^{9}-10^{12}$ can 
easily be accommodated, thus reconciling thermal leptogenesis with the gravitino problem. 
In Sec.~\ref{sec:conclude}, we present a summary and conclusions.

\section{Mixed axion/axino relic density}
\label{sec:Oh2}

In this section, we list the four production mechanisms
for mixed axion/axino dark matter which are considered here.
It is possible that other more exotic mechanisms could also contribute, 
such as axino production from moduli or inflaton decay. 
We will not consider these additional mechanisms here.

\subsection{Axions via vacuum misalignment}

Here, we consider the scenario where the PQ symmetry breaks before the 
end of inflation, so that a nearly uniform value of the axion field
$\theta_i\equiv a(x)/(f_a/N)$ is expected throughout the universe.
The axion field equation of motion implies that the axion field stays relatively
constant until temperatures approach the QCD scale $T_{QCD}\sim 1$ GeV.
At this point, a temperature-dependent axion mass term turns on, and a 
potential is induced for the axion field. 
The axion field rolls towards its minimum and oscillates, 
filling the universe with low energy (cold) axions. 
The expected axion relic density via this vacuum mis-alignment mechanism is 
given by~\cite{vacmis}
\be
\Omega_a h^2\simeq 0.23 f(\theta_i)\theta_i^2 
\left(\frac{f_a/N}{10^{12}\ {\rm GeV}}\right)^{7/6}
\ee
where $0< \theta_i<\pi$ and $f(\theta_i)$ is the so-called anharmonicity
factor. Visinelli and Gondolo~\cite{vacmis} parametrize the latter as
$f(\theta_i)=\left[\ln\left(\frac{e}{1-\theta_i^2/\pi^2}\right)\right]^{7/6}$.
The uncertainty in $\Omega_a h^2$ from vacuum mis-alignment is estimated 
as plus-or-minus a factor of three. 

\subsection{Thermal production of axinos}

If the reheat temperature $T_R$ exceeds the axino decoupling temperature
\be
T_{dcp}=10^{11}\ {\rm GeV}\left(\frac{f_a/N}{10^{12}\ {\rm GeV}}\right)^2
\left(\frac{0.1}{\alpha_s}\right)^3 ,
\ee
axinos will be in thermal equilibrium, 
with an abundance given by $\Omega_{\ta}^{TE}h^2\simeq\frac{m_{\ta}}{2\ {\rm keV}}$. 
To avoid overproducing axino dark matter, the RTW bound~\cite{rtw} 
then implies that $m_{\ta}<0.2$ keV.
We will here consider only reheat temperatures below $T_{dcp}$. 
In this case, the axinos are never in thermal equilibrium in the early universe.
However, they can still be produced thermally via radiation off of particles that are
in thermal equilibrium~\cite{ckkr,steffen}. Here, we adopt a recent calculation of the thermally produced
axino abundance from Strumia~\cite{strumia}:
\be
\Omega_{\ta}^{\rm TP}h^2=1.24 g_3^4 F(g_3)\frac{m_{\ta}}{{\rm GeV}}
\frac{T_R}{10^4\ {\rm GeV}}\left(\frac{10^{11}}{f_a/N}\right)^2 ,
\ee
with $F(g_3)\sim 20 g_3^2\ln\frac{3}{g_3}$, and $g_3$ is the strong coupling constant
evaluated at $Q=T_R$.

\subsection{Axinos via neutralino decay} 

In supersymmetric scenarios with the neutralino as a quasi-stable NLSP, the 
$\tz_1$s will be present in thermal equilibrium in the early universe, and will
freeze out when the expansion rate exceeds their interaction rate, at a temperature
roughly $T_{f}\sim m_{\tz_1}/20$. 
The present day abundance can be evaluated by integrating the Boltzmann equation. 
Several computer codes are available for this computation. Here we use the
code IsaReD~\cite{isared}, 
a part of the Isajet/Isatools package~\cite{isatools,isajet}.

In our case, each neutralino will undergo decay to the stable axino LSP, via decays
such as $\tz_1\to\ta\gamma$. Thus, these non-thermally produced axinos will
inherit the thermally produced neutralino number density, and we will simply have~\cite{ckkr}
\be
\Omega_{\ta}^{\tz}h^2=\frac{m_{\ta}}{m_{\tz_1}}\Omega_{\tz_1}^{TP}h^2  .
\ee

\subsection{Axinos from gravitino cascade decay}

Since here we are attempting to generate reheat temperatures $T_R\agt 10^9$ GeV, 
we must also include in our calculations the thermal production of gravitinos in the
early universe. 
We here follow Pradler and Steffen, who  have estimated the thermal gravitino production
abundance as~\cite{relic_G}
\be
\Omega_{\tG}^{\rm TP}h^2 =\sum_{i=1}^{3}\omega_ig_i^2(T_R)
\left(1+\frac{M_i^2(T_R)}{3m_{\tG}^2}\right)\ln\left(\frac{k_i}{g_i(T_R)}\right)
\left(\frac{m_{\tG}}{100\ {\rm GeV}}\right)\left(\frac{T_R}{10^{10}\ {\rm GeV}}\right) ,
\ee
where $\omega_i=(0.018,0.044,0.117)$, $k_i=(1.266,1.312,1.271)$, $g_i$ are the gauge 
couplings evaluated at $Q=T_R$ and $M_i$ are the gaugino masses also evaluated at $Q=T_R$.
For the temperatures we are interested in, this agrees
with the calculation by
Rychkov and Strumia\cite{relic_G} within a factor of about 2.

Since each gravitino cascade decays ultimately to an axino LSP, the
abundance of axinos from gravitino production is given by
\be
\Omega_{\ta}^{\tG}h^2=\frac{m_{\ta}}{m_{\tG}}\Omega_{\tG}^{TP}h^2 .
\ee
For axino masses in the MeV range and gravitino masses in the $30-50$~TeV range,
the prefactor above is extremely small, and allows us to evade overproduction of
dark matter via thermal gravitino production (part 2 of the gravitino problem).
\footnote{
We have checked that for gravitinos in the mass range $5\ {\rm TeV}<m_{\tG}<50$ TeV, 
the temperature $T_{decay}=\sqrt{\Gamma_{\tG}m_{Pl}}/(\pi^2g_*/90)^{1/4}$ at which
gravitinos decay ranges
between 0.01-0.5 MeV, well after neutralino freeze-out. This means that the gravitino 
cascade decays indeed contribute to the ultimate axino relic density, rather than having the neutralinos
from the cascade reprocessed in the thermal bath, as would occur for $T_{decay}\agt m_{\tz_1}/20$. 
}
\footnote{
Along with gravitino cascade decays, 
the decay $\tG\to a\ta$ can also occur, and give rise to a component of hot dark matter
axions. Since this contribution to the axion density is $\sim (m_a/m_{\tG})\Omega_{\tG}h^2$ 
with $m_a\sim \mu{\rm eV}$, it will be a
very tiny contribution to the total dark matter density, and so we are safe to neglect it here.
}

\subsection{Mixed axion/axino dark matter}

In this paper, we will evaluate the mixed axion/axino relic density from the above four sources:
\be
\Omega_{a\ta}h^2=\Omega_{a}h^2+\Omega_{\ta}^{\rm TP}h^2+
\Omega_{\ta}^{\tz}h^2+\Omega_{\ta}^{\tG}h^2 .
\label{eq:ata}
\ee
Over much of parameter space, if $m_{\ta}$ is taken to be of order the MeV scale, then the
contributions from $\tz_1$ and gravitino production are subdominant. 
In Fig.~\ref{fig:Oh2}, we illustrate in the upper frame the relative importance 
of the four individual contributions as a function of $f_a/N$, 
for an mSUGRA scenario with $m_{\tG}=1$ TeV, $m_{\tz_1}=122$ GeV, $\Omega_{\tz_1}h^2=9.6$ 
(as in Ref. \cite{axdm}).
For the axion/axino sector we take $\theta_i=0.05$ and $m_{\ta}=100$ keV. 
The value of $T_R$ is adjusted such that $\Omega_{a\ta}h^2=0.1123$. 
For low $f_a/N$ values, the TP axino contribution is dominant. But as $f_a/N$ increases, 
the axion component grows until at $f_a/N\sim 4\times 10^{13}$ GeV it becomes dominant, 
and for even higher $f_a/N$ it saturates the DM relic density. 

\FIGURE[t]{
\includegraphics[width=10cm]{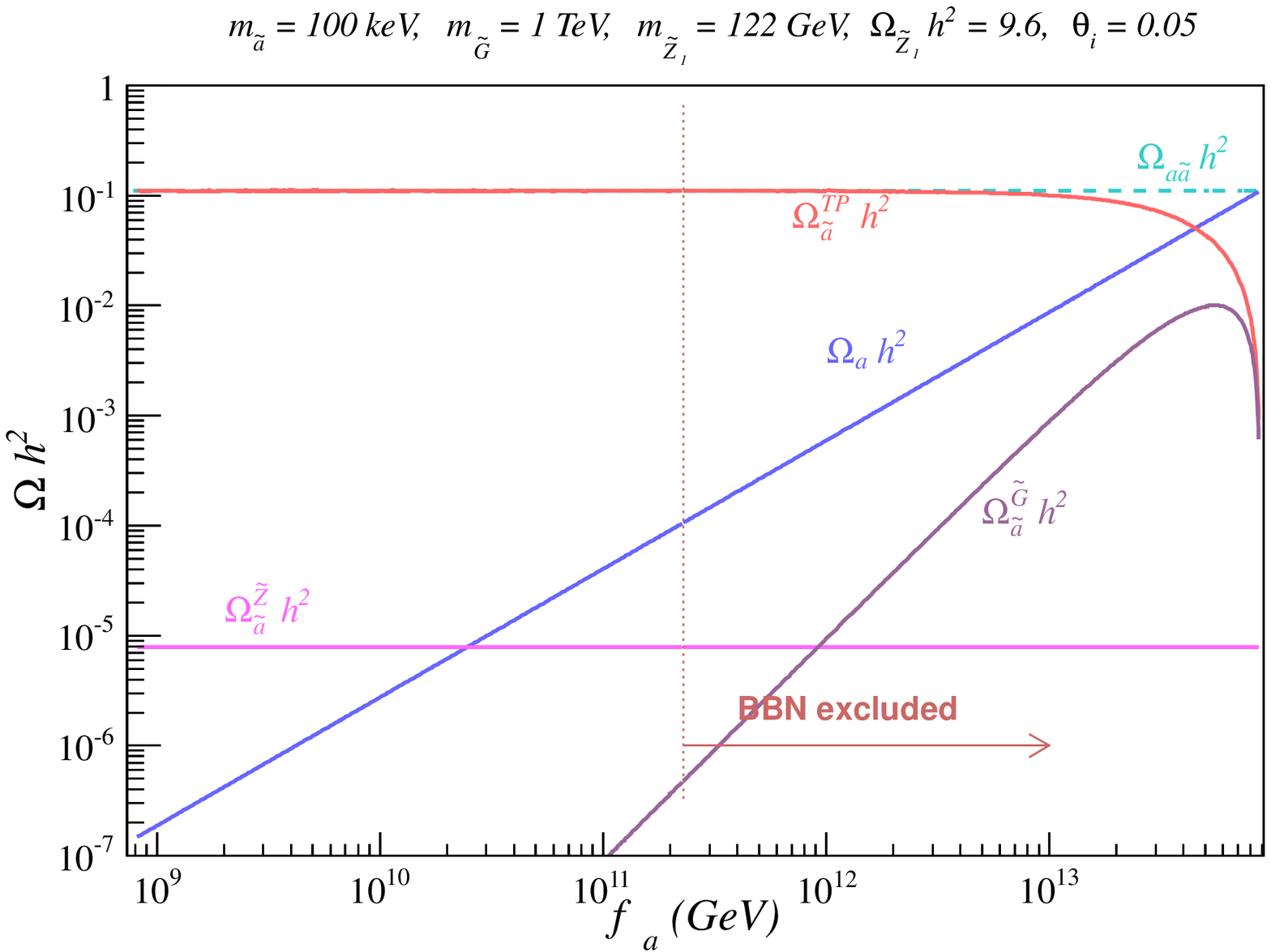}
\includegraphics[width=10cm]{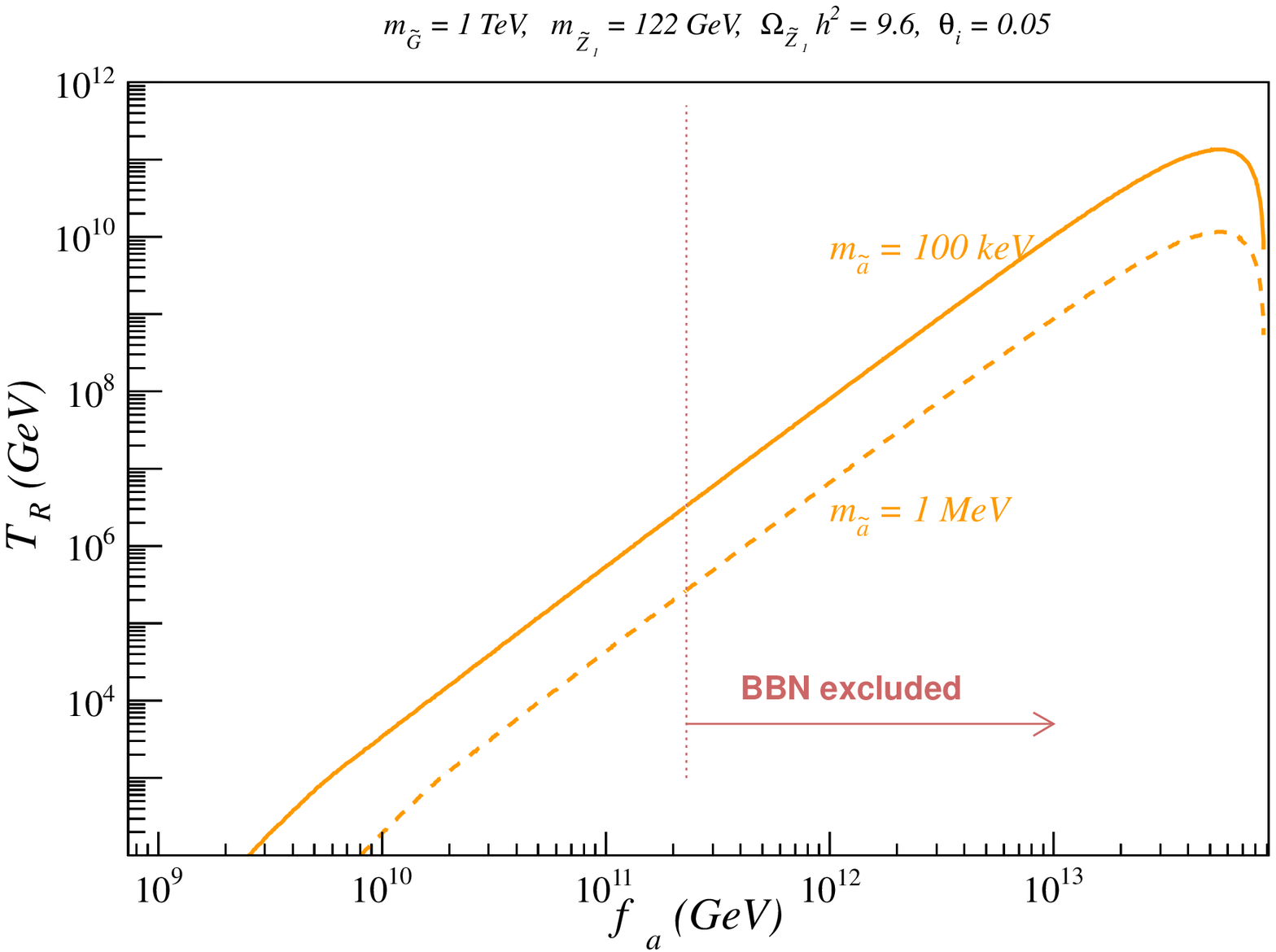}
\caption{Upper frame: Contribution of axions and TP and NTP  
axinos to the DM density as a function of the PQ breaking scale $f_a/N$, 
for an mSUGRA point with $m_0=1000$ GeV, $m_{1/2}=300$ GeV, $A_0=0$, $\tan\beta =10$ and
$\mu >0$, and fixing $m_{\ta}=100$ keV and $\theta_i=0.05$; $T_R$ is adjusted such that 
$\Omega_{a\ta}h^2=0.1123$. 
Lower frame: the $T_R$ that is needed to achieve $\Omega_{a\ta} h^2 = 0.1123$ 
for $m_{\ta} = 0.1$ and 1 MeV, for the same mSUGRA point and $\theta_i$.
}\label{fig:Oh2}}

The value of $T_R$ which is needed is shown in the lower frame of Fig.~\ref{fig:Oh2}. 
We see that $T_R$ grows quickly with increasing $f_a/N$. This is because the thermal 
axino production decreases as the inverse square of $f_a/N$, so larger values of 
$T_R$ are needed to keep $\Omega_{a\ta}h^2 = 0.1123$.
We see that $T_R$ can reach $\sim 10^{11}$ GeV in the case of mainly axion CDM (similar to
Ref.~\cite{axdm}). In our case here, allowing a smaller value of $\theta_i$ 
allows higher values of $f_a/N$ to be found, which in turn requires much higher 
values of $T_R$, into the range needed for thermal leptogenesis. 

However, for such high $f_a/N$, the $\tz_1$ becomes so long-lived that it violates 
the bounds from BBN on late decaying neutral particles (as indicated in the figure). 
We address this issue in the next section.

\section{Neutralino lifetime and hadronic branching fraction}
\label{sec:z1}

We have averted one problem with BBN by requiring the presence of a gravitino with 
$m_{\tG}\agt 30$ TeV, so that it will decay largely before BBN starts.
In the process, by asking for
$T_R\agt 2\times 10^9$ GeV while avoiding overproduction of mixed axino/axion dark matter
(the latter requires large $f_a/N\sim 10^{12}$ GeV and small $\theta_i$), 
we have pushed the $\tz_1$ lifetime uncomfortably high, so that its hadronic decays in the 
early universe now have the potential to disrupt BBN.

Constraints from BBN on hadronic decays of long-lived neutral particles in the early universe
have been calculated by several authors~\cite{ellis,kohri,jedamzik}. Here, we will adopt the results
from the recent calculations by Jedamzik~\cite{jedamzik}. The BBN constraints arise due to injection of 
high energy hadronic particles into the thermal plasma during or after BBN. The constraints depend
on three main factors:
\bi
\item The abundance of the long-lived neutral particles. In Ref.~\cite{jedamzik}, this
is given by $\Omega_X h^2$ where $X$ is the long-lived neutral particle which undergoes hadronic decays.
In our case, where the long-lived particle is the lightest neutralino which decays to an axino LSP,
this is just given by the usual thermal neutralino abundance $\Omega_{\tz_1}h^2$, as calculated
by IsaReD~\cite{isared}. 
\item The lifetime $\tau_X$ of the long-lived neutral particle. Obviously, the longer-lived
$X$ is, the greater its potential to disrupt the successful BBN calculations.
\item The hadronic branching fraction $B_h$ of the long-lived neutral particle. If is very small,  
then very little hadronic energy will be injected, and hence the constraints should be more mild.
\ei
The BBN constraints are shown in Fig.~9 (for $m_X=1$ TeV) and Fig.~10 (for $m_X=100$ GeV)
of Ref.~\cite{jedamzik}, as contours in the $\tau_X\ vs.\ \Omega_Xh^2$ plane, with numerous
contours for differing $B_h$ values ranging from $10^{-5}$ to 1. 
For $B_h\sim 0.1$, for instance, and very large values
of $\Omega_Xh^2\sim 10-10^3$, the lifetime $\tau_X$ must be $\alt 0.1$~sec, or else the
primordial abundance of $^4He$ is disrupted. 
If $\Omega_X h^2$ drops below $\sim 1$, then much larger
values of $\tau_X$ up to $\sim 100$ sec are allowed. If one desires a long-lived hadronically
decaying particle in the early universe with $\tau_X\agt 100$ sec, then typically much lower
values of $\Omega_X h^2\sim 10^{-6}-10^{-4}$ are required. Such low neutralino relic densities
are extremely hard to generate in SUSY models, even in the case of AMSB, or pure higgsino
annihilation~\cite{ax19}. We have digitized the constraints of Ref.~\cite{jedamzik}, implementing
extrapolations for cases intermediate between values of parameters shown, so as to
approximately apply the BBN constraints to our scenario with a long-lived neutralino
decaying hadronically during BBN.

\subsection{Neutralino lifetime}

We have calculated the two-body decays of $\tz_1$ to axinos, and find agreement with the results
of Ref.~\cite{ckkr}. In the notation of Ref.~\cite{wss}, we find
\be
\Gamma (\tz_1\to\ta\gamma )=\frac{\alpha_Y^2C_{aYY}^2}{128\pi^3}
\frac{v_4^{(1)2}\cos^2\theta_W}{(f_a/N)^2}
m_{\tz_1}^3\left(1-\frac{m_{\ta}^2}{m_{\tz_1}^2}\right)^3
\ee
and
\bea
\Gamma (\tz_1\to\ta Z )& =&\frac{\alpha_Y^2C_{aYY}^2}{128\pi^3}\frac{v_4^{(1)2}\sin^2\theta_W}{(f_a/N)^2}
m_{\tz_1}^3
\lambda^{1\over 2}(1,\frac{m_Z^2}{m_{\tz_1}^2},\frac{m_{\ta}^2}{m_{\tz_1}^2})\nonumber  \\
& &\left[(1+\frac{m_{\ta}}{m_{\tz_1}})^2-\frac{m_Z^2}{m_{\tz_1}^2}\right]
\left[(1-\frac{m_{\ta}}{m_{\tz_1}})^2+\frac{m_Z^2}{2m_{\tz_1}^2}\right] \,,
\eea
with $\alpha_Y=(e^2/4\pi)/\cos^2\theta_W$ the $U(1)_Y$ coupling;  
$C_{aYY}=8/3$ in the DFSZ model~\cite{dfsz} and 0, $2/3$ or $8/3$ in the KSVZ model~\cite{ksvz}
depending on the heavy quark charge $e_Q=0$, $-1/3$ or $2/3$. Throughout our analysis
we assume $C_{aYY}=8/3$. The above decays should be the only two-body decay modes allowed for the KSVZ model;
for the DFSZ model, additional decays to higgs states may also be allowed. 
Since such decays dependent on the type of DFSZ model, we do not consider them in our analysis.

Using the above formulae, in Fig.~\ref{fig:tau} we plot the lifetime
$\tau (\tz_1 )$ in seconds versus $m_{\tz_1}$ for various choices of
$(f_a/N)/v_4^{(1)}$. The quantity $v_4^{(1)}$ denotes the bino-component of $\tz_1$
in the notation of Ref.~\cite{wss}. For models with a bino-like $\tz_1$, 
$v_4^{(1)}\sim 1$. For models with a wino-like $\tz_1$, as in AMSB, the 
lifetime will be enhanced by a large factor, since in these models $v_4^{(1)}$
is typically $10^{-2}-10^{-3}$.

\FIGURE[t]{
\includegraphics[width=9cm]{lifetime.eps}
\caption{Lifetime (in seconds) of a bino-like $\tz_1$ with a $\ta$ 
as LSP versus $m_{\tz_1}$, for various choices of $(f_a/N)/v_4^{(1)}$ (in GeV units).
We take $C_{aYY}=8/3$.
}\label{fig:tau}}

From Fig.~\ref{fig:tau}, we see that for models with a bino-like neutralino and
$\tau (\tz_1 )\alt 0.01$ sec, either very small $f_a/N\alt 10^{10}$ GeV are required, or if
larger $f_a/N$ values are desired, then $m_{\tz_1}$ must be very (perhaps uncomfortably) large.
However, if $\tau (\tz_1 )<10^2$ sec is needed, then values of $f_a/N$ as large
as $10^{14}$ GeV are allowed, depending on $m_{\tz_1}$.

\subsection{Two- and three-body $\tz_1$ decay to hadrons}

Finally, to implement BBN constraints, we will need the hadronic branching fraction of
$\tz_1$ decay. If the decay $\tz_1\to\ta Z$ is open, then $B_h$ is just given by
$B_h=\frac{\Gamma (\tz_1\to \ta Z)}{\Gamma_{\tz_1}}\times BF(Z\to hadrons)$. When the decay 
$\tz_1\to \ta Z$ is closed, we must instead calculate the three-body decay 
$\tz_1\to \ta q\bar{q}$. This decay is calculated in Ref.~\cite{ckkr} via 
$\gamma$ and $Z^*$ exchange diagrams, but neglecting interference terms. 
Here, we present the three-body width including interference:
\bea
\frac{d\Gamma}{d\mu_k}&=&\frac{m_{\tz_1}^3}{12\pi^3}\left[G_\gamma^2e^2Q^2\frac{(1-\mu_k)^2(2+\mu_k)}{\mu_k}
+g_Z^2G_Z^2(g_V^2+g_A^2)\frac{(1-\mu_k)^2(2+\mu_k)\mu_k}{\frac{m_Z^2\Gamma_Z^2}{m_{\tz_1}^4}+(\frac{m_Z^2}{m_{\tz_1}^2}-\mu_k)^2}\right.\nonumber \\
& & +\left. 2g_ZQeg_V Re(G_\gamma^*G_Z)Re\left[\frac{(1-\mu_k)^2(2+\mu_k)}{\mu_k-\frac{m_Z^2}{m_{\tz_1}^2}
+i\frac{\Gamma_Zm_Z}{m_{\tz_1}^2}}\right]\right] ,
\eea
where the axino and quark masses have been neglected.
In the above, $G_Z=\frac{\alpha_YC_{aYY}}{16\pi}\frac{v_4^{(1)}}{f_a/N}\sin\theta_W$,
$G_\gamma =\frac{\alpha_YC_{aYY}}{16\pi}\frac{v_4^{(1)}}{f_a/N}\cos\theta_W$, 
$g_Z=\frac{e}{\sin\theta_W\cos\theta_W}$, $g_V={T_3\over 2}-Q\sin^2\theta_W$ and 
$g_A=-T_3/2$, where $T_3$ is the weak isospin of the quark $q$. The above 
differential width is integrated over the range 
$\mu_k: 4m_q^2/m_{\tz_1}^2 \to 1$.
The quark mass acts as a regulator for the otherwise divergent photon-mediated contribution.

The hadronic branching fraction $B_h$ of $\tz_1$ is plotted in Fig.~\ref{fig:BFZ1} 
versus $m_{\tz_1}$. The values of $v_4^{(1)}$, $C_{aYY}$ and $f_a/N$ cancel out in 
the branching fraction calculation, so the result is quite general. We see that for 
low values of $m_{\tz_1}\alt m_Z$, $B_h\sim 0.02$. 
Once $m_{\tz_1}$ exceeds $m_Z$, the branching fraction for $\tz_1\to\ta Z$ turns on 
and $B_h$ increases asymptotically towards $\sim 0.175$.
Armed with the values of $\Omega_{\tz_1}h^2$, $\tau (\tz_1 )$, $B_h$ and the constraints 
of Ref.~\cite{jedamzik}, we are now ready to explore allowed regions of PQMSSM parameter
space which might reconcile thermal leptogenesis with the gravitino problem and at the
same time satisfy the BBN contraints for late decaying neutralinos.

\FIGURE[t]{
\includegraphics[width=9cm]{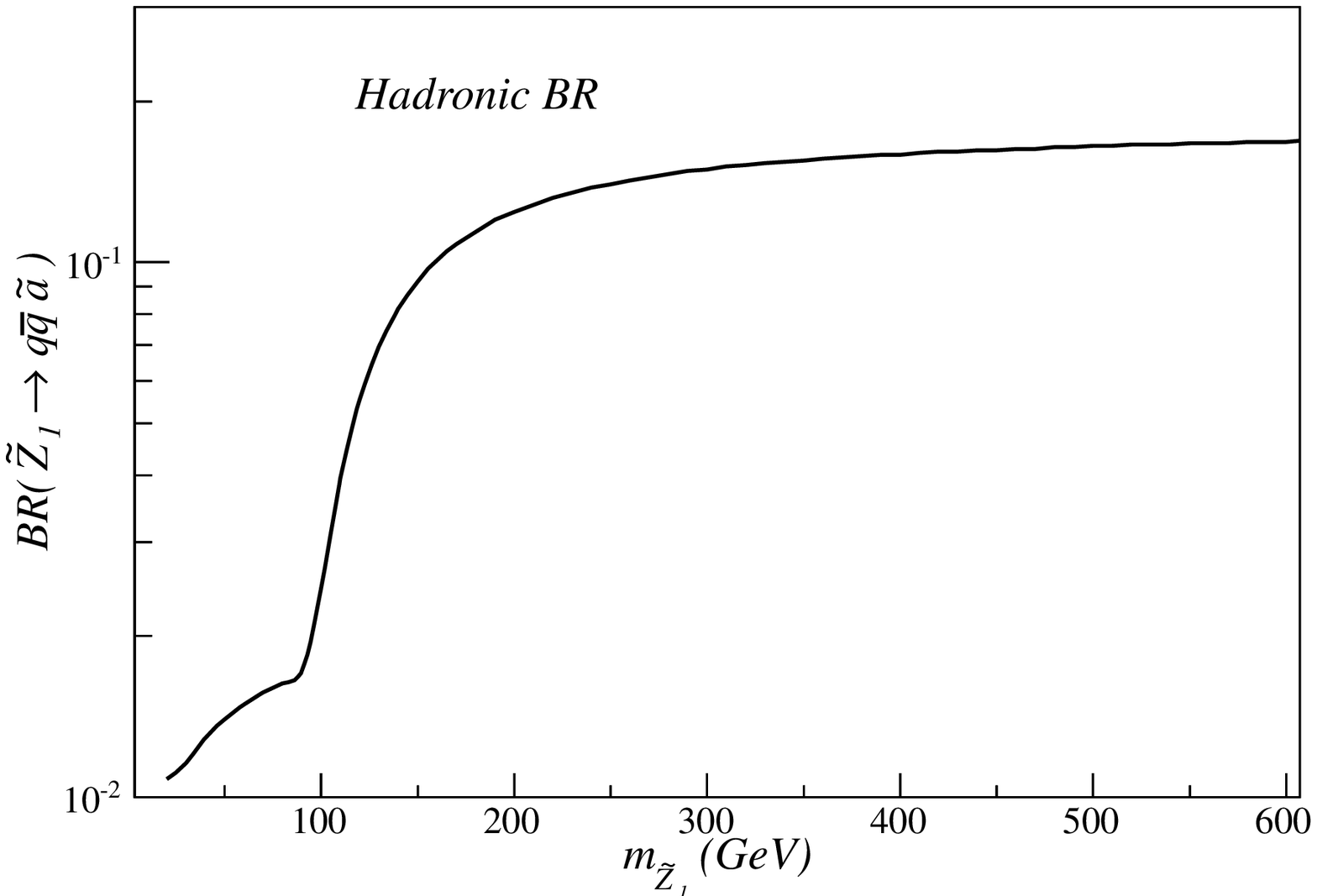}
\caption{Branching fraction of $\tz_1\to\ta +hadrons$ versus $m_{\tz_1}$.
}\label{fig:BFZ1}}

\section{Allowed values of $T_R$ in supersymmetric models}
\label{sec:TR}

\subsection{Five benchmark models}

In this section, we discuss the sorts of models which might allow
a reconciliation of thermal leptogenesis with the gravitino problem.
Our starting point is to require models with a rather heavy gravitino
$m_{\tG}\agt 30$ TeV, so as to avoid the gravitino BBN constraint. We also invoke
mixed axion/axino dark matter with a light $\ta$ ($m_{\ta} \lesssim$ 1 MeV) so as
to avoid the gravitino-induced overclosure problem.

Models with multi-TeV scale gravitinos are rather limited.
In ordinary gravity-mediation (SUGRA) models, the gravitino mass
arises from the superHiggs mechanism in a hidden sector of the model.
The gravitino mass then sets the scale for the soft SUSY breaking terms.
This latter condition applies to the scalar sector in simple SUGRA models,
while gaugino masses require in addition stipulation of
the gauge kinetic function. We will assume here that gauginos are quite light;
if instead gaugino masses are in the multi-TeV range, then RGE effects
drive the third generation scalar soft masses into the multi-TeV range as well, thus
engendering a conflict with naturalness.

Two types of SUGRA models 
have multi-TeV (1st and 2nd generation) scalar masses, while the 3rd generation 
is around the TeV scale, as motivated by the hierarchy problem. 
The first is Yukawa-unified SUSY (YU), in which scalar masses are preferred
to be in the multi-TeV range at the GUT scale, while gauginos need to
be quite light. In these models, the large, unified 
third generation Yukawa couplings drive the third generation soft terms 
down into the TeV range while first and second generation soft terms
at $Q=M_{weak}$ remain in the multi-TeV regime~\cite{hb,bdr,bkss}. 
The SUSY particle mass spectrum for $\mu>0$ is characterized as a radiatively 
driven inverted scalar mass hierarchy~\cite{imh}. When requiring consistency 
with B-physics constraints, the lightest neutralino 
is a nearly pure bino state, while scalars are quite heavy, thus suppressing
the neutralino annihilation cross section. 
Generally, the neutralino relic density is large, of order
$\Omega_{\tz_1}h^2\sim 10-10^4$, far beyond the measured value.
By invoking instead a light axino as LSP, the relic density 
of mixed axion/axino dark matter can be reconciled with observation~\cite{bs,bhkss}.
We present in Table~\ref{tab:bm} a Yukawa-unified benchmark model
(model HSb from Ref.~\cite{lhc7}) with $m_{16}=10$ TeV, which would be consitent 
with gravity mediation with a $\sim 30$ TeV gravitino mass. For this point, 
the lightest neutralino has mass $m_{\tz_1}=49$ GeV, and $\Omega_{\tz_1}h^2=3195$.

\begin{table}\centering
\begin{tabular}{lccccc}
\hline
 & BM1  & BM2  & BM3  & BM4  & BM5  \\
 &  (YU) & (ESUSY) &  (inoAMSB) &  (MU) &  (MU) \\
\hline
$m_{3/2}$ [TeV]  & 30 & 30 & 50 & 30 & 30 \\
$m_{16}\ {\rm or}\ m_0$   & 10000 & 20575.6  & 0 & -- &  -- \\
$m_{16}(3)$   & -- & 2922.94  & -- & -- &  -- \\
$m_{1/2}\ {\rm or}\ M_2$  & 43.94 & 1457.17 & 161.4 & -- & -- \\
$A_0$      & $-19947.3$ & 2177.84 & 0 & -- & -- \\
$m_{H_d}$   & 12918.9 & 3099.42 & -- & -- & -- \\
$m_{H_u}$   & 11121.0 & 2783.53 & -- & -- & -- \\
$\tan\beta$  & 50.398 & 6.87475 & 10 & 10 & 10 \\
$\alpha$ & -- & -- & -- & $-1.6$ & $6$ \\
$n_m,\ n_H$ & -- & -- & -- & $0,\ 0$ & ${1\over 2},\ 0$ \\
\hline
$\mu$      & 3132.6 & 418.6 & 598.6 & 1136.8 & 992.6 \\
$m_{\tG}$   & 351.2 & 3507.1 & 1129.7 & 1354.1 & 1903.9 \\
$m_{\tu_L}$ & 9972.1 & 20739.8 & 993.9 & 1327.0 & 1770.1 \\
$m_{\tst_1}$& 2756.5 & 652.8 & 861.6 & 804.3 & 1040.8 \\
$m_{\tb_1}$ & 3377.1 & 671.7 & 926.2 & 1123.7 & 1517.4 \\
$m_{\te_L}$ & 9940.7 & 20613.9 & 229.4 & 433.1 & 983.0 \\
$m_{\tw_1}$ & 116.4 & 428.0 & 142.4 & 164.9 & 945.4 \\
$m_{\tz_2}$ & 113.8 & 425.3 & 443.5 & 164.6 & 943.7 \\ 
$m_{\tz_1}$ & 49.2 &  414.2 & 142.1 & 146.0 & 759.2 \\ 
$m_A$       & 1825.9 &  2832.5 & 632.8 & 1190.2 & 1584.1 \\
$m_h$       & 127.8 &  117.5 & 112.1 & 117.4 & 121.3 \\ \hline
$\Delta a_\mu$ & $5.9\times 10^{-12}$ & $2.4\times 10^{-13}$ & $1.6\times 10^{-9}$ 
& $-1.5\times 10^{-10}$ & $1.3\times 10^{-10}$\\
$BF(b\to s\gamma )$ & $3.1\times 10^{-4}$ & $2.9\times 10^{-4}$ & $3.8\times 10^{-4}$ 
& $3.5\times 10^{-4}$ & $3.0\times 10^{-4}$\\
$BF(B_s\to\mu\mu)$ & $8.9\times 10^{-9}$ & $3.8\times 10^{-9}$ & $3.8\times 10^{-9}$ & 
$3.8\times 10^{-9}$ & $3.9\times 10^{-9}$ \\
$v_4^{(1)}$ & 1 & 0.14 & 0.009 & 1 & $-0.99$ \\
$\Omega h^2_{\tz_1}$ & 3195 & 0.04 & 0.0016 & 0.04 & 0.06 \\
$\sigma (\tz_1 p)$ [pb] & $3.3\times 10^{-13}$  & $6.6\times 10^{-9}$ & $4.4\times 10^{-9}$ & $3.1\times 10^{-11}$ & 
$1.4\times 10^{-9}$ \\
\hline
\end{tabular}
\caption{Masses and parameters in~GeV units for the five benchmark points. 
BM\,2--5 are computed with \Isajet\,7.81 using $m_t=173.1$ GeV.
BM1 uses Isajet 7.79 with $m_t=172.6$ to be consistent with previous work. 
}
\label{tab:bm}
\end{table}

The second type of gravity mediation model which would be consistent with
$\sim 30$ TeV gravitinos is effective SUSY, or ESUSY~\cite{ckn}. 
In Ref.~\cite{esusy}, these models were explored with GUT scale soft SUSY
breaking boundary conditions. Viable spectra with a weak scale inverted scalar
mass hierarchy were found. For ESUSY models,  
first/second generation scalars could have mass $m_0(1,2)$ in the multi-TeV 
range at the GUT scale, while third generation scalar masses $m_0(3)$ are in 
the few TeV range.
Upon RG evolution, first/second generation scalars remain in the multi-TeV range, 
while third generation scalar masses are suppressed by two-loop RGE terms,
and are sub-TeV at the weak scale. Benchmark point BM2 is an example of an
ESUSY scenario.

If we proceed beyond gravity-mediation, then the class of models which necessarily 
supports multi-TeV gravitinos is anomaly-mediation (AMSB)~\cite{amsb}. In these models,
sparticle masses arise at the loop level via the superconformal anomaly.
Sparticle masses are of order $m\sim \frac{g_i^2}{16\pi^2}m_{3/2}$, so that in order
to support TeV scale sparticle masses, a gravitino mass of order $50-200$~TeV is needed.
In AMSB models, the lightest neutralino is nearly a pure wino state with a relic density 
usually well below the measured $\Omega_{\rm DM}h^2$, unless $m_{\tz_1}\agt 1300$~GeV.
For benchmark model BM3, we select a gaugino AMSB (inoAMSB) point with 
$m_{3/2}=50$~TeV.\footnote{We use different notations for the gravitino mass scale 
$m_{3/2}$ and the physical gravitino mass $m_{\tG}$, though $m_{\tG}\approx m_{3/2}$.} 
It has been argued in Ref.~\cite{shanta} that in string models the scalar soft masses 
and trilinear terms are actually suppressed, while gaugino masses assume the usual AMSB form.
These models, with $m_0=A_0=0$ at $M_{GUT}$, avoid the problem of tachyonic scalars
which occurs in traditional AMSB models; the scalar masses are uplifted via RG running
during their trajectories from $M_{GUT}$ to $M_{weak}$. While we do select the inoAMSB model
as our benchmark point, very similar dark matter phenomenology occurs for
minimal AMSB (mAMSB) or hypercharged AMSB~\cite{hcamsb}, since the defining characteristics
are a wino-like lightest neutralino~\cite{shibi}.

The third class of models we examine also easily supports multi-TeV gravitinos. 
These are the mixed moduli-AMSB models~\cite{mmamsb}, 
also known as mirage unification (MU).
This class of models is inspired by the KKLT set-up of string models with flux 
compactifications and an uplifted scalar potential which can accommodate a positive 
cosmological constant~\cite{kklt}. While MU models
require a multi-TeV gravitino mass, the lightest neutralino can easily remain bino-like, 
and can also have a very low relic abundance $\Omega_{\tz_1}h^2$ at the 0.1 level or below.
These models are stipulated by the parameters $\alpha$, which governs how much gravity versus
anomaly mediation occurs, along with $m_{3/2}$ and $\tan\beta$. One must also
stipulate the matter and Higgs field modular weights $n_m$ and $n_H$, which take on values of 
$0$, $1\over 2$ or 1. The MU models are hard coded into the Isasugra spectrum generator.

Benchmark model BM4 takes $\alpha =-1.6$, $m_{3/2}=30$ TeV, $\tan\beta =10$ and $n_m=n_H=0$.
It has a bino-like lightest neutralino with $m_{\tz_1}=146$ GeV, but $\Omega_{\tz_1}h^2=0.04$ due to 
bino-wino co-annihilation, or BWCA~\cite{bwca}. In BWCA, the gaugino masses $M_1\simeq -M_2$
at the weak scale. Since the gaugino masses have opposite signs, there is no mixing between
bino and wino states, although they can be close in mass and can thus co-annihilate.
If instead $M_1\simeq +M_2$ at the weak scale, then one obtains a $\tz_1$ of
mixed bino-wino content (which also occurs in MU models).

Benchmark model BM5 is also of the MU type, but with $\alpha =6$, $m_{3/2}=30$ TeV,
$\tan\beta =10$ and $n_m={1\over 2}$, $n_H=0$. This model yields a bino-like $\tz_1$
with mass $m_{\tz_1}=759$ Gev, but $\Omega_{\tz_1}h^2=0.06$ due to neutralino
annihilation through the pseudoscalar $A$-resonance~\cite{Afunnel}.\\

In the following, we examine whether these scenarios are compatible with a $T_R$ 
high enough to allow for thermal leptogenesis. To this aim we perform for each 
BM point a random scan over PQMSSM parameters in the range
\bea
 m_{\ta}  &\in& [10^{-7},\;10^{-1}]\ {\rm GeV}\,,\nonumber \\
 f_a/N    &\in& [10^8,\;10^{14}]\ {\rm GeV}\,, \label{eq:scan}\\
 \theta_i &\in& [0,\;\pi]\,. \nonumber
\eea
and calculate the value of $T_R$ which is needed to enforce $\Omega_{a\ta}=0.1123$.
As mentioned, a major constraint 
comes from the BBN bounds on the lifetime of the $\tz_1$. 
A digitized version of these BBN bounds is shown in Fig.~\ref{fig:BBN}, 
in the $\tau (\tz_1)$ versus $\Omega_{\tz_1}h^2$ plane.
We also show the locus of benchmark points BM1--BM5 on the plot, along with the
respective maximum allowed values of $f_a/N$ which are consistent with the BBN bounds.

\FIGURE[t]{
\includegraphics[width=10cm]{BBNbounds3.eps}
\caption{BBN bounds on late-decaying neutral particles with
$B_h=0.1$, digitized from Ref.~\cite{jedamzik}, in the
$\tau (\tz_1)$ versus $\Omega_{\tz_1}h^2$ plane.
We also show the locus of benchmark points BM1--BM5, along with
their maximum allowed $f_a/N$ values.
}\label{fig:BBN}}

\subsection{Gravity mediation: Yukawa unified SUSY}

The YU model point BM1 has $m_{\tz_1}=49$ GeV, $v_4^{(1)}=1$ and 
$\Omega_{\tz_1}= 3195$. In addition we take $m_{\tG}=30$ TeV. 
Since $\Omega_{\tz_1}h^2$ is so large, the $\tz_1$ lifetime is restricted
to be $\alt 0.03$ sec by the BBN bounds from Ref.~\cite{jedamzik}, 
see Fig.~\ref{fig:BBN}. The neutralino lifetime bound then translates into 
an upper bound on $f_a/N\alt 3\times 10^{10}$~GeV. 

\FIGURE[t]{
\includegraphics[width=10cm]{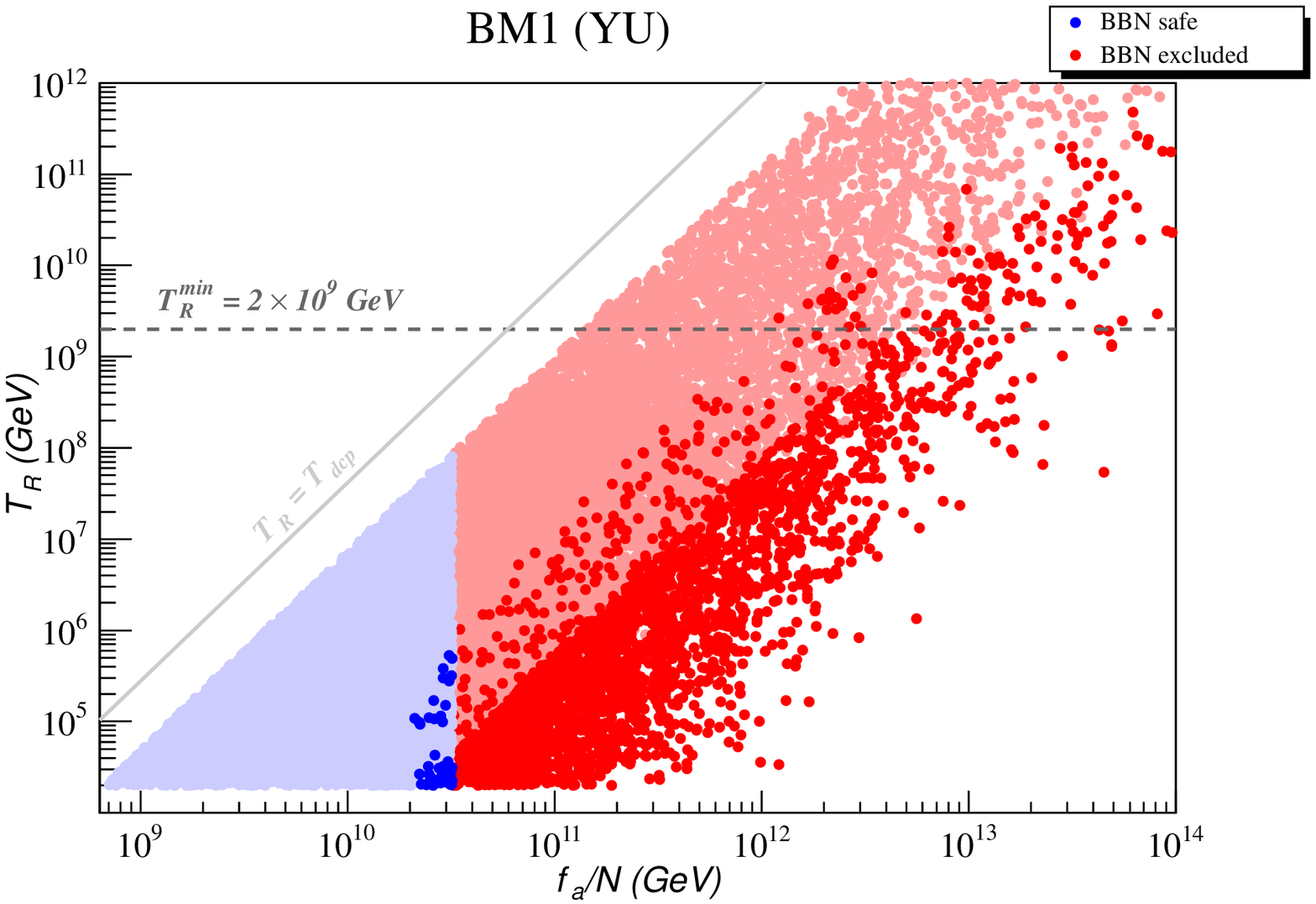}
\caption{Scan over PQMSSM parameters for BM1 (YU model) plotted in the $T_R\ vs.\ f_a/N$ plane. 
The blue points respect the $\tz_1\to\ta+hadrons$ BBN bound, while the red points 
violate the BBN constraint. Points shown in light blue or light red have 
$>20\%$ WDM or $>1\%$ HDM, as discussed in the text.
}\label{fig:BM1fa}}

The results of the scan over PQMSSM parameters, Eq.~(\ref{eq:scan}), are shown in 
Fig.~\ref{fig:BM1fa} in the $f_a/N\ vs.\ T_R$ plane. 
For $T_R$ values above the diagonal line labelled $T_R=T_{dcp}$, the axinos would have
been in thermal equilibrium in the early universe; all points lie below, so the expression 
for $\Omega_{a\ta}h^2$ in Eq.~(\ref{eq:ata}) is valid.
In the figure, the [light and dark] blue dots denote points consistent with the 
neutralino BBN bound, while [light and dark] red dots denote points which violate BBN 
constraints. As shown in Fig.~\ref{fig:Oh2}, the largest $T_R$ values are mostly obtained 
for a light axino. Depending on its mass, the axino might constitute 
warm (WDM) or hot (HDM) dark matter, which is severely constrained by the matter 
power spectrum and reionization~\cite{ckkr,jlm}, see also~\cite{warm,hot}.
Since the bounds on the amount of HDM/WDM are model dependent~\cite{jlm}, we do 
not impose such constraints on our results. However, as a guidance, we indicate by 
lighter colors the points which have:
\bi
\item $m_{\ta} < 100$ keV and $\Omega_{\ta}/\Omega_{a\ta} > 0.2$ or 
\item $m_{\ta} < 1$ keV and $\Omega_{\ta}/\Omega_{a\ta} > 0.01$, 
\ei
where $\Omega_{\ta} = \Omega_{\ta}^{TP} + \Omega_{\ta}^{\tG} + \Omega_{\ta}^{\tz}$. 
Dark blue and dark red points thus have mostly CDM with at most 20\%~WDM and 
1\%~HDM admixture.\footnote{A rough estimate based on the neutrino mass 
limit~\cite{hot} from cosmological data, $\sum m_\nu<0.41$ to $0.44$~eV, 
gives that up to 4--5\% HDM contribution could be acceptable.} 

As can be seen in Fig.~\ref{fig:BM1fa}, for values of $f_a/N$ consistent with BBN bounds,
only $T_R$ values below $10^8$ GeV are allowed.
Such low values of $T_R$ are insufficient to support thermal leptogenesis, but are
sufficient to support non-thermal leptogenesis, which requires a more modest value of
$T_R\agt 10^6$~GeV~\cite{bs,bhkss}. Note also that these points 
(with $T_R\sim 10^6-10^8$~GeV) have a substantial fraction of WDM. 
In fact, as we will see later, for $\tz_1$ masses of about 50~GeV, 
as typical for YU scenarios with $\mu>0$, reconciling thermal leptogenesis 
with the gravitino problem requires a very low neutralino abundance, 
cf.\ Fig.~\ref{fig:Oh2vsFA}. Such low abundances
can indeed be achieved in a small region of the YU SUSY parameter
space \cite{bkss} or--perhaps more easily--using either generational
non-universality, or gaugino mass non-universality \cite{auto}.

\subsection{Gravity mediation: Effective SUSY (ESUSY)}

Another gravity-mediation model which requires multi-TeV scalars is  
effective SUSY. The ESUSY point BM2 has a mixed bino-higgsino NLSP with mass 
$m_{\tz_1}=414$~GeV. The $\tz_2$ and $\tw_1$ are quite close in mass to the $\tz_1$, 
followed by $\tilde t_1$ and $\tilde b_1$ which are just 60\% heavier. 
This leads to $\Omega_{\tz_1}h^2\sim 0.04$ due to simultaneous mixed bino-higgsino-wino 
enhanced annihilation, and also a contribution from stop and sbottom co-annihilation. 
The low $\Omega_{\tz_1}h^2$ value allows $\tz_1$
lifetimes up to $\sim 200$~sec, corresponding to $f_a/N$ values as high as $10^{13}$ GeV
(see Fig.~\ref{fig:BBN}). 

Such high $f_a/N$ values suppress the thermal production of axino dark matter, while
low values of $\theta_i$ suppress the axion relic abundance.
Scanning over PQMSSM parameters reveals that re-heat temperatures above $10^{12}$~GeV
can be generated while avoiding overproduction of dark matter and maintaining consistency
with BBN bounds. Thus, ESUSY models with a low abundance of neutralinos, mixed axion/axino 
dark matter with a high PQ scale and low $\theta_i$, 
apparently can reconcile thermal leptogenesis with the gravitino problem, although 
most of the solutions for BM2 have a potentially dangerous fraction of HDM/WDM.

We point out, however, that the BBN bounds will be less severe for similar 
scenarios with larger $\mu$ ({\it i.e.}\ less $\tz_1$ higgsino admixture) but lighter 
$\tilde t_1$ or $\tilde b_1$, such that a low $\Omega_{\tz_1}h^2$ arises from 
stop or sbottom co-annihilation; examples are discussed in~\cite{esusy}.

\FIGURE[t]{
\includegraphics[width=10cm]{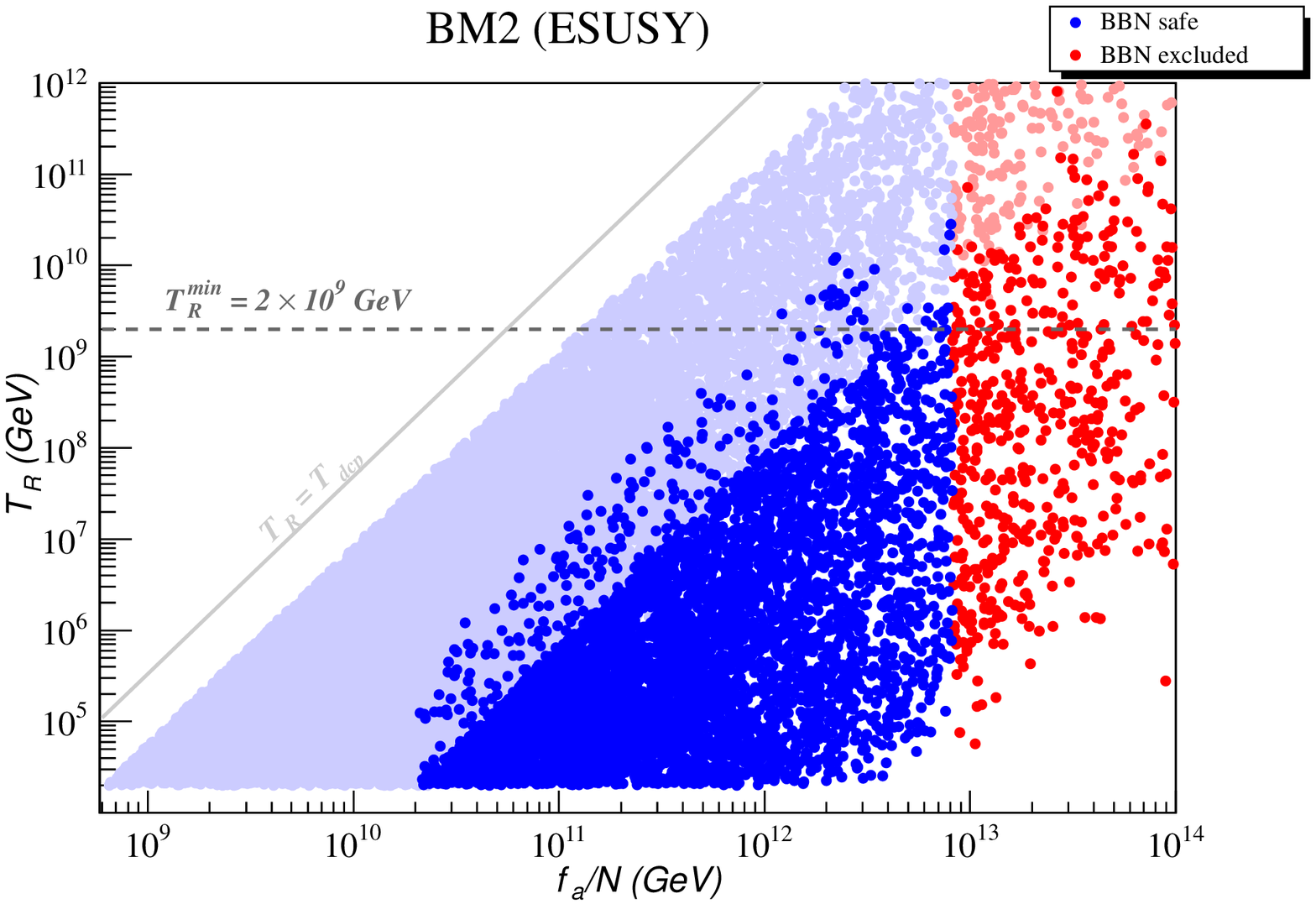}
\caption{Scan over PQMSSM parameters for BM2 (ESUSY model) plotted in the 
$T_R\ vs.\ f_a/N$ plane. Same color code as in Fig.~\ref{fig:BM1fa}.
}\label{fig:BM2fa}}

\subsection{Anomaly mediation: gaugino AMSB}

Next, we consider a inoAMSB model, point BM3. In this case, 
the relic abundance is much lower, $\Omega_{\tz_1}h^2=0.016$, thus much longer
$\tz_1$ lifetimes of $\sim 800$ sec are allowed. Nevertheless, the $\tz_1$ 
lifetime is suppressed in this case by the tiny value of $v_4^{(1)}\sim 0.01$, 
so that for BM3, $f_a/N$ values only as high as $2\times 10^{11}$~GeV are 
allowed. 
The results are shown in Fig.~\ref{fig:BM3fa}, where we again show BBN-allowed 
and BBN-forbidden model points in the $f_a/N\ vs.\ T_R$ plane. We see that just 
a few points barely exceed the rough requirement for thermal leptogenesis 
that $T_R>2\times 10^9$~GeV.
If we increase $m_{\tG}$ beyond 50 TeV, then the value of $\Omega_{\tz_1}h^2$ 
increases, requiring shorter $\tz_1$ lifetimes, although the $\tz_1$ lifetime 
also decreases as $\sim 1/m_{\tz_1}^3$. 
We also note that the light blue points with $T_R>2\times 10^{9}$~GeV 
have axino masses of a few times $10^{-7}$~GeV. 
Overall, we conclude that AMSB models with a wino-like $\tz_1$ can just 
barely reconcile thermal leptogenesis with the gravitino problem, however 
this would require a considerable fraction of axino HDM.

\FIGURE[t]{
\includegraphics[width=10cm]{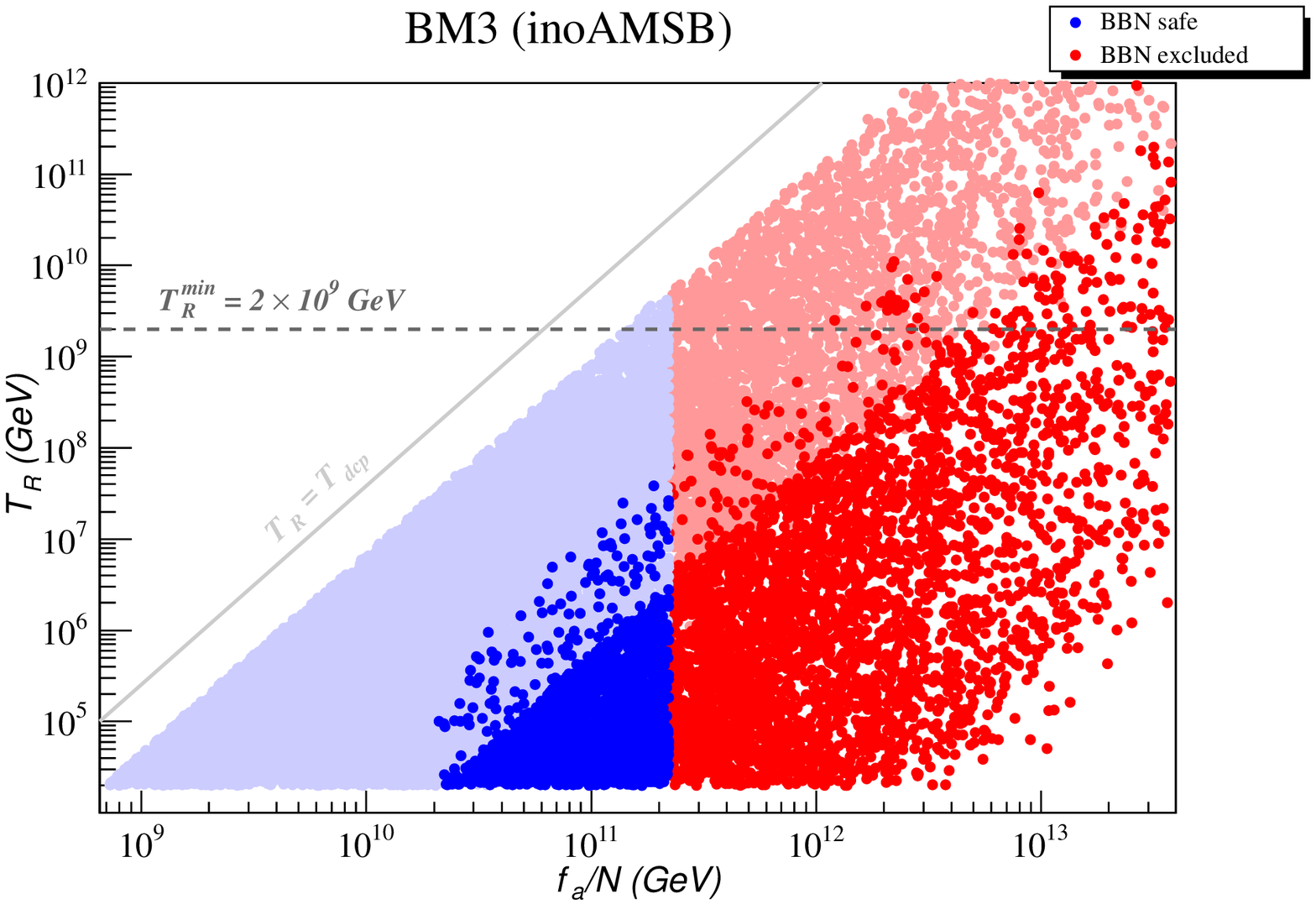}
\caption{Scan over PQMSSM parameters for BM3 (gaugino AMSB model) plotted 
in the $T_R\ vs.\ f_a/N$ plane. Same color code as in Fig.~\ref{fig:BM1fa}.
}\label{fig:BM3fa}}

\subsection{Mixed moduli/anomaly mediation with bino-wino co-annihilation}

Let us now move to mirage unification models, and the BM4 point with bino-wino 
co-annihilation. In this case, the neutralino mass is $146$ GeV and its relic 
density $\Omega_{\tz_1}h^2=0.04$, so that $\tz_1$ lifetimes of $\sim 200$~sec 
are allowed. Since the $\tz_1$ is nearly 
pure bino, its decay is unsuppressed by $v_4^{(1)}$, and we find 
in Fig.~\ref{fig:BM4fa} that $f_a/N$ values as high as $10^{13}$~GeV are allowed
by BBN constraints. With such high $f_a/N$ values, the thermal production of
axinos is suppressed, and $T_R$ values over $10^{12}$~GeV are allowed. Thus, these
models are capable of reconciling thermal leptogenesis with the gravitino problem
while avoiding BBN constraints.

In Fig.~\ref{fig:BM4th}, we plot the scanned points for the BM4 point in the
$\theta_i\ vs.\ T_R$ plane. (In fact the analogous plot for BM2 looks almost 
the same.) We see that the BBN-allowed blue points
must have $\theta_i$ on the small side, certainly $<1$ in order to avoid
overproducing axions via vacuum mis-alignment, while maintaining
$T_R\agt 10^9$ GeV.

Figure~\ref{fig:BM4max} shows the scan results for BM4 in the 
$m_{\ta}\ vs.\ T_R$ plane. Here, we see that the points which are BBN-allowed, 
and also are consistent with reconciling thermal leptogenesis with the gravitino 
problem require $m_{\ta}\alt 100$ keV. For values of $m_{\ta}$ lower than 100~keV, 
the axinos may start becoming warm rather than cold dark matter, 
and for values lower than 1~keV they contribute to HDM. Thus,
if CDM/WDM constraints are properly applied, we expect that a small region of 
parameter space will be consistent with thermal leptogenesis, as roughly 
indicated by the dark blue points in Figs.~\ref{fig:BM4fa}--\ref{fig:BM4max}. 
%
%

\FIGURE[t]{
\includegraphics[width=10cm]{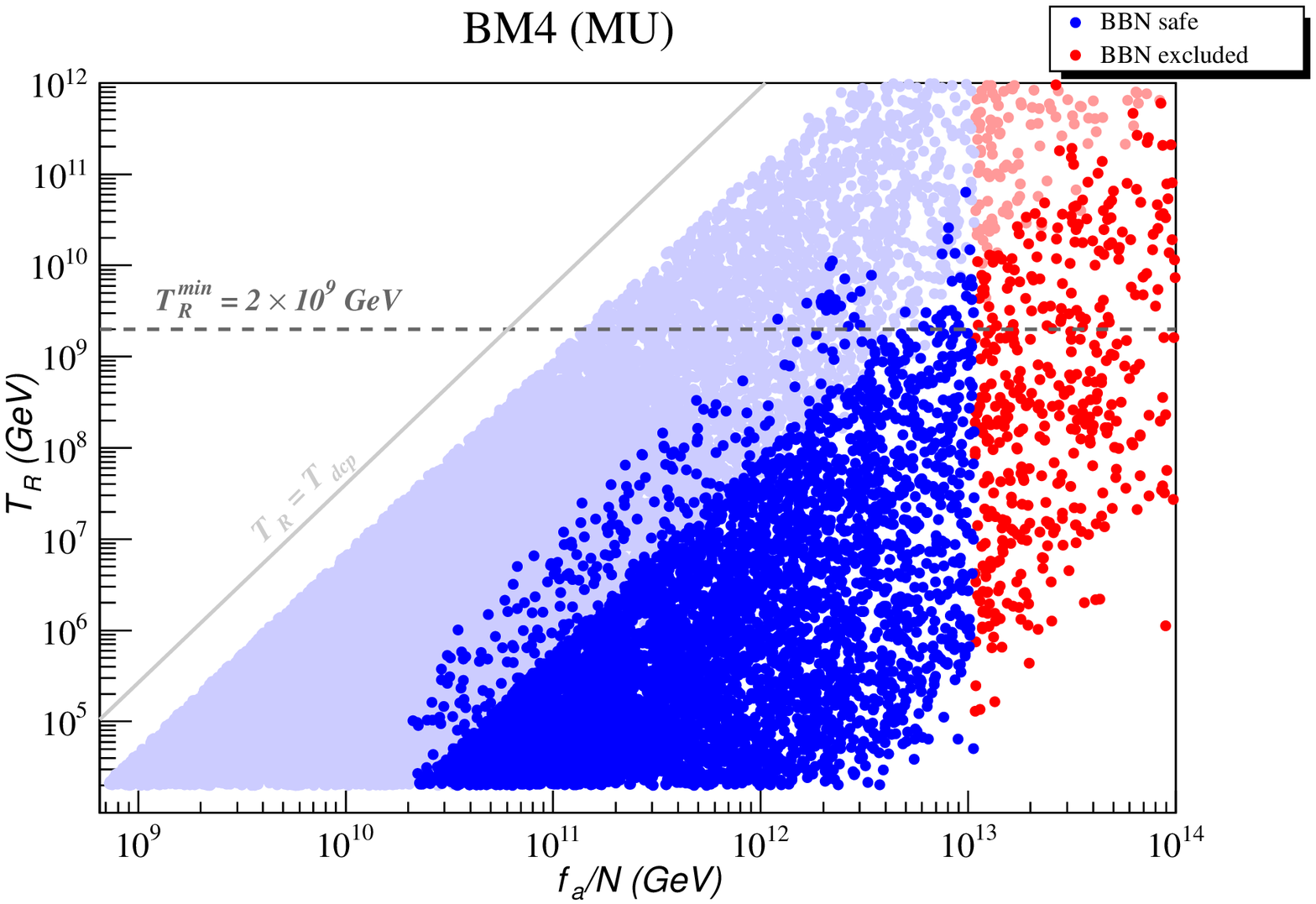}
\caption{Scan over PQMSSM parameters for BM4 (MM-AMSB with bino-wino co-annihilation) 
plotted in the $T_R\ vs.\ f_a/N$ plane. Same color code as in Fig.~\ref{fig:BM1fa}.
}\label{fig:BM4fa}}

\FIGURE[t]{
\includegraphics[width=10cm]{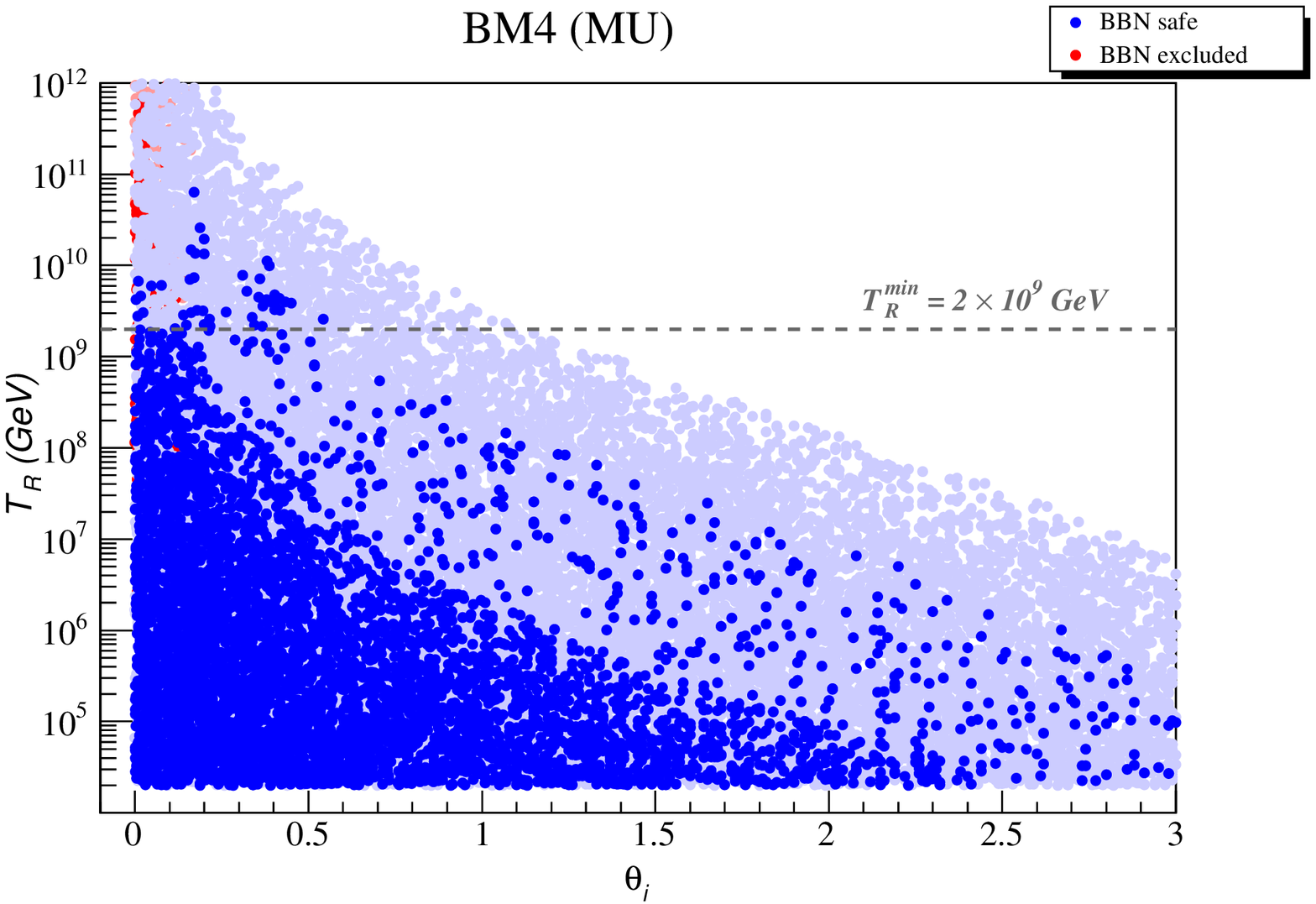}
\caption{Scan over PQMSSM parameters for BM4 plotted in the $T_R\ vs.\ \theta_i$ 
plane. Same color code as in Fig.~\ref{fig:BM1fa}.
}\label{fig:BM4th}}

\FIGURE[t]{
\includegraphics[width=10cm]{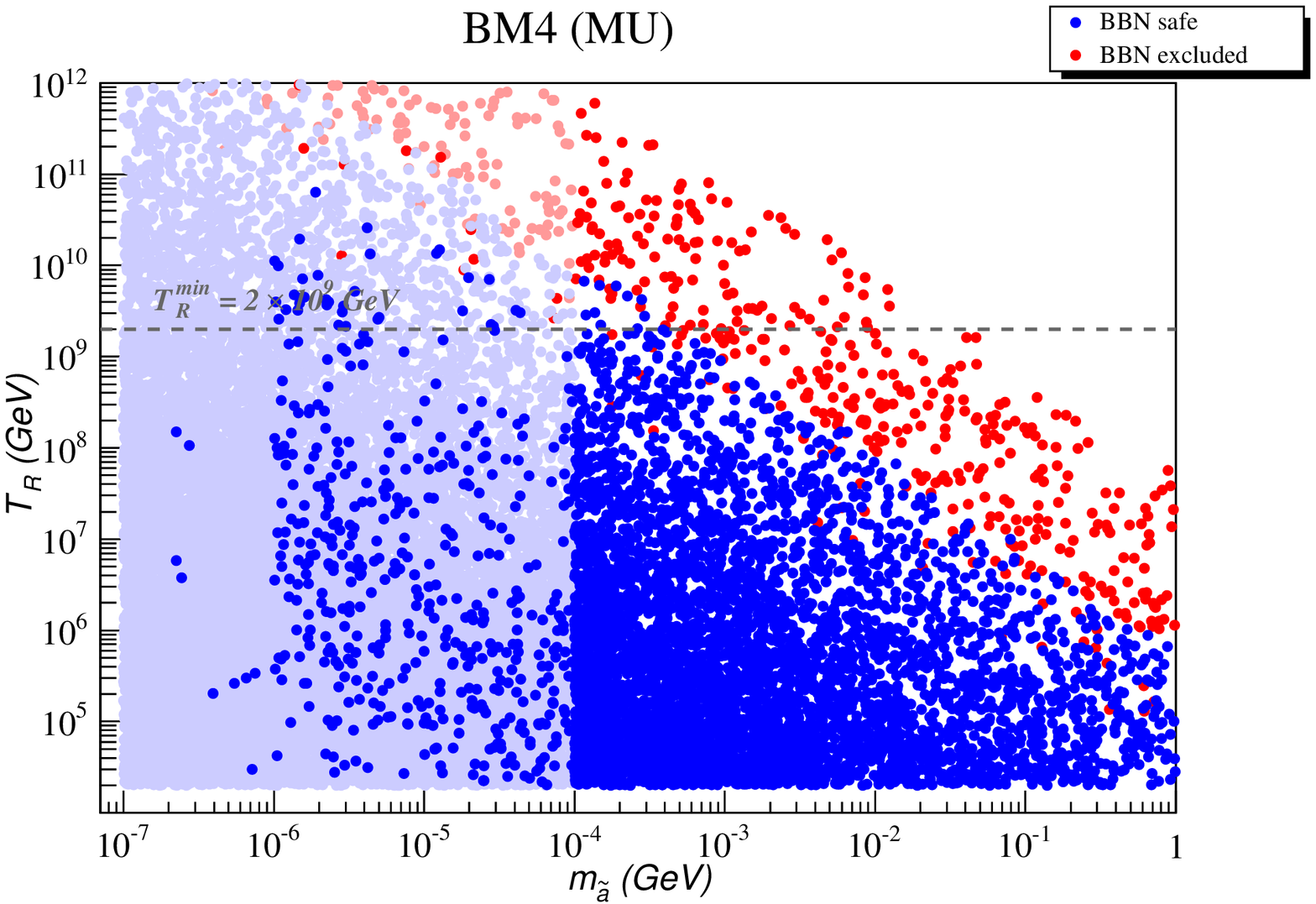}
\caption{Scan over PQMSSM parameters for BM4 plotted in the $T_R\ vs.\ m_{\ta}$ 
plane. Same color code as in Fig.~\ref{fig:BM1fa}.
}\label{fig:BM4max}}

\subsection{Mixed moduli/anomaly mediation with neutralino annihilation
on the $A$-resonance}

Next, we turn to benchmark point BM5, a MU model with a bino-like neutralino 
with $m_{\tz_1}=759$ GeV and $\Omega_{\tz_1}h^2=0.06$ due to 
neutralino annihilation through the $A$-resonance ($m_A=1584$ GeV, so that
$2m_{\tz_1}\sim m_A$). The low value of $\Omega_{\tz_1}h^2$ again allows
for $\tz_1$ lifetimes as high as $\sim 200$~sec. But now, since $m_{\tz_1}$
is so large, $f_a/N$ values up to $\sim 10^{14}$ GeV are allowed by Fig.~\ref{fig:tau}.
The scanned points are shown in Fig.~\ref{fig:BM5fa}. We see that all points 
pass the BBN constraints, allowing for $T_R$ values in excess of $10^{12}$ GeV. 
This model again easily reconciles thermal leptogenesis with the gravitino problem.

As in the case of BM4, $\theta_i$ values must be $\alt 1$, and for very high 
$T_R>10^{11}$ GeV, $\theta_i<0.4$. 
We conclude that small values of $\theta_i$ are necessary to allow
for thermal leptogenesis in SUSY models with mixed axion/axino dark matter.
We also show in Fig.~\ref{fig:BM5max} the scan result for BM5 in the
$m_{\ta}\ vs.\ T_R$ plane. Here we see that consistency with thermal leptogenesis 
requires axino masses $m_{\ta}\alt 10$ MeV.
Thermally produced axinos with mass $\agt 0.1$ MeV should constitute cold dark matter,
so these points would have a mix of cold axions plus cold thermally produced axinos.

\FIGURE[t]{
\includegraphics[width=10cm]{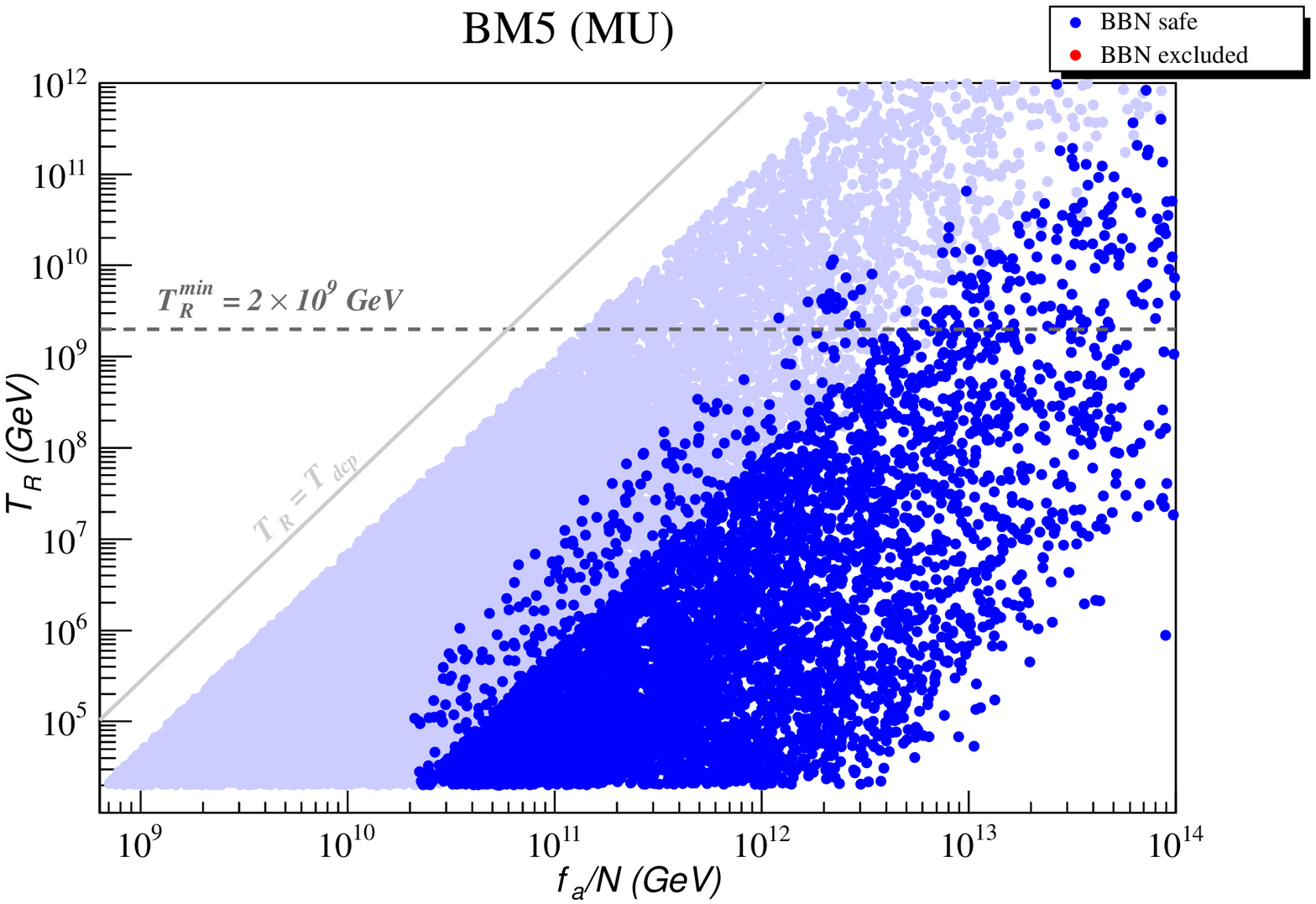}
\caption{Scan over PQMSSM parameters for BM5 (MU with $A$-resonance annihilation) 
plotted in the $T_R\ vs.\ f_a/N$ plane. Same color code as in Fig.~\ref{fig:BM1fa}.
}\label{fig:BM5fa}}

\FIGURE[t]{
\includegraphics[width=10cm]{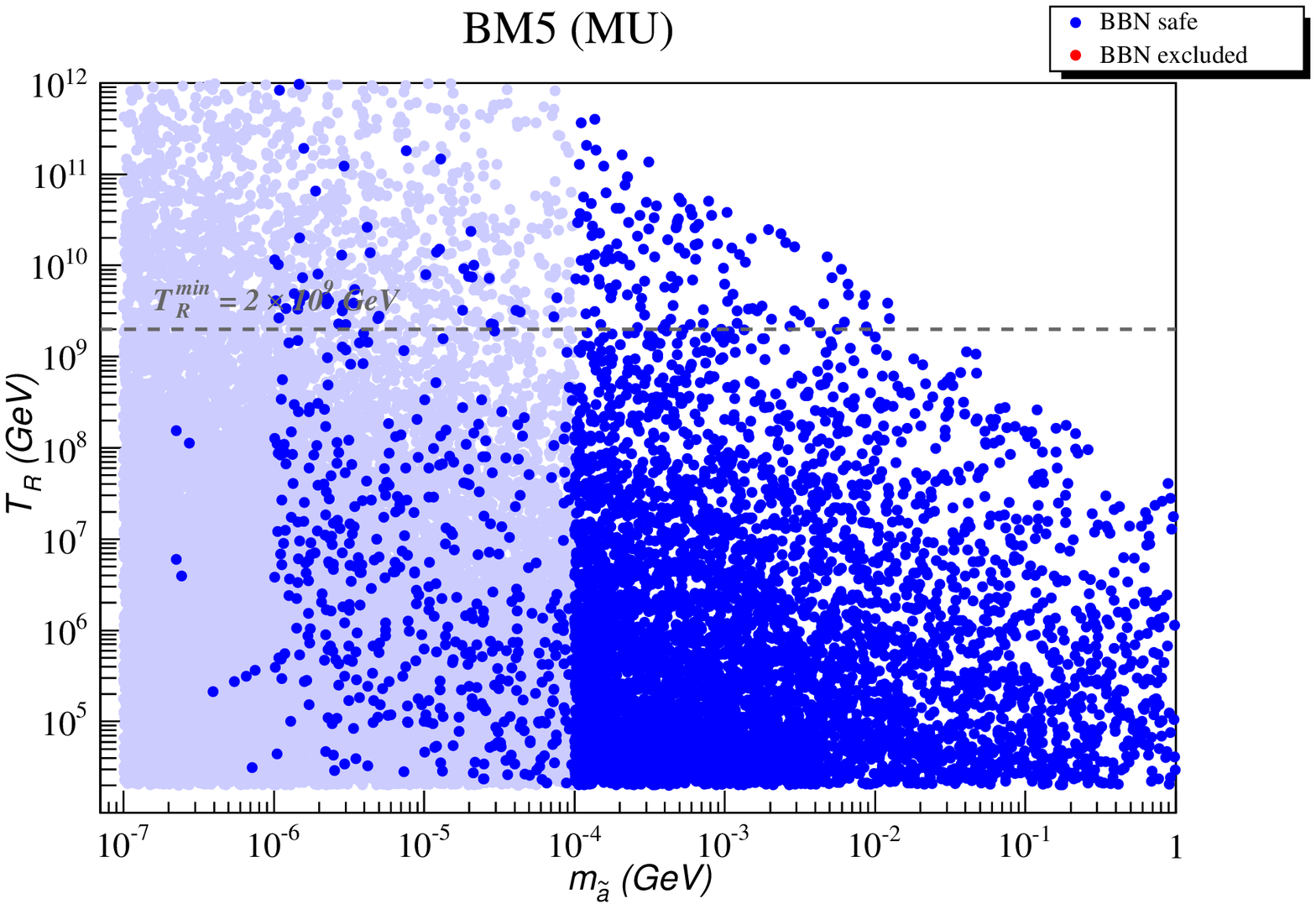}
\caption{Scan over PQMSSM parameters for BM5 MU model plotted in the $T_R\ vs.\ m_{\ta}$ 
plane. Same color code as in Fig.~\ref{fig:BM1fa}.
}\label{fig:BM5max}}

\section{Conclusions}
\label{sec:conclude}

In this paper, we examined $R$-parity conserving supersymmetric 
models with a goal of reconciling thermal leptogenesis with the 
gravitino problem, via the postulation of mixed axion/axino dark matter.
The mixed dark matter arises naturally from the Peccei-Quinn solution
to the strong $CP$ problem in supersymmetric models.

In order to reconciliate thermal leptogenesis with cosmological
constraints, such as the dark matter relic abundance and the BBN bounds
on late decaying particles, we conclude that the following conditions are
necessary:
\bi
\item $T_R \gtrsim 2\times10^9$~GeV, to allow for efficient
thermal production of right-handed neutrinos in the early universe;
\item $m_{\tG}\agt 30$ TeV,
to avoid BBN constraints on gravitino production in the early universe.
Models with such a heavy gravitino are also favored in SUGRA models in that
they set the scale for the scalar soft mass terms; if the scalar masses
are sufficiently high, then they can suppress unwanted FCNC  and CP
violating processes and also proton decay via a decoupling solution~\cite{dine};
\item An axino LSP in keV to MeV mass range; this condition allows us to avoid 
overproduction of dark matter from axino thermal production in the early universe 
as well as from $\tz_1$ and $\tG$ decays, 
since the matter density is then suppressed by the ratio $m_{\ta}/m_{\tz_1,\tG}$. 
We also require a neutralino NLSP, which is common in 
supersymmetric models with heavy scalar masses;
\item $f_a/N\agt 10^{12}$ GeV, to suppress thermal overproduction of axino dark matter.
The high value of $f_a/N$ means the axion mass is likely
to lie in the sub-micro-eV range;
\item $\theta_i\lesssim 1$, to avoid overproduction of axions when $f_a/N > 10^{12}$ GeV;
\item $\Omega_{\tz_1} h^2 \lesssim 1$, $m_{\tz_1} \gtrsim 100$ GeV and/or $v_4^{(1)} \sim 1$, 
to avoid BBN constraints on the late decaying neutralino.
\ei

For the SUGRA case, we examined Yukawa-unified and effective SUSY
models, since these can easily accommodate a 30~TeV gravitino mass.
These models typically have too low an annihilation cross section 
for the neutralino NLSP in the early universe, which leads to conflicts with
BBN constraints on late decaying neutral particles: in this case, hadronic
neutralino decay via $\tz_1 \to Z^*/\gamma \to \ta q\bar{q}$.

The ESUSY model can more easily allow for low neutralino abundances 
via stop, sbottom, stau or higgsino co-annihilation.
In these models, requiring the sum of four production mechanisms for
mixed axion/axino DM to equal the measured abundance can allow for
$T_R\agt 10^{10}$~GeV, thus reconciling thermal leptogenesis with 
the gravitino problem.

In AMSB models, one naturally has a gravitino mass in the 50--100~TeV range 
and small $\Omega_{\tz_1} h^2$, but with a wino-like neutralino.
The decay rate of the
wino-like $\tz_1$ is suppressed by the mixing factor $v_4^{(1)\,2}$, leading to
long-lived $\tz_1$s, and likely conflicts with BBN.

We also examined models with mixed moduli-AMSB (mirage unification)
soft terms. These models allow for 30--100 TeV gravitinos, but with 
a bino-like neutralino, so its lifetime is typically $\alt 100$~sec.
We examined two cases: bino-wino co-annihilation and $A$-resonance
annihilation. Both cases easily allow $T_R$ to reach over 
$10^{12}$ GeV, thus easily reconciling thermal leptogenesis with the
gravitino problem, while respecting BBN constraints on 
long-lived neutralinos.

These findings are summarized in a model-independent way in Fig.~\ref{fig:Oh2vsFA}, 
which shows PQMSSM scan points with $T_R \gtrsim 2\times10^9$~GeV in 
the $\Omega_{\tz_1} h^2\ vs.\ f_a/N$ plane for $m_{\tz_1}=50$ and 
500~GeV.\footnote{Additional entropy production from saxion decay may 
soften the BBN bounds; this is left for future work.}
In both cases, $v_4^{(1)}=1$; recall that the neutralino lifetime scales as 
$[v_4^{(1)}/(f_a/N)]^2$.
It is interesting to note that the reconciliation of thermal
leptogenesis and the gravitino problem, in the framework used for our
analysis, spans a wide range of $\Omega_{\tz_1} h^2$, from 
$\Omega_{\tz_1} h^2\sim 1$ down to very low values. In particular for 
a light neutralino NLSP, very low $\Omega_{\tz_1} h^2$ is required.
Models with neutralino DM, on the other hand, would have 
$\Omega_{\tz_1} h^2 \sim 0.1$, while models with a light axino DM and 
non-thermal leptogenesis would prefer $\Omega_{\tz_1} h^2 \gtrsim 100$.
Therefore distinct regions of the (PQ)MSSM parameter space are
prefered depending on which DM and baryogenesis solutions are chosen.

\FIGURE[t]{
\includegraphics[width=10cm]{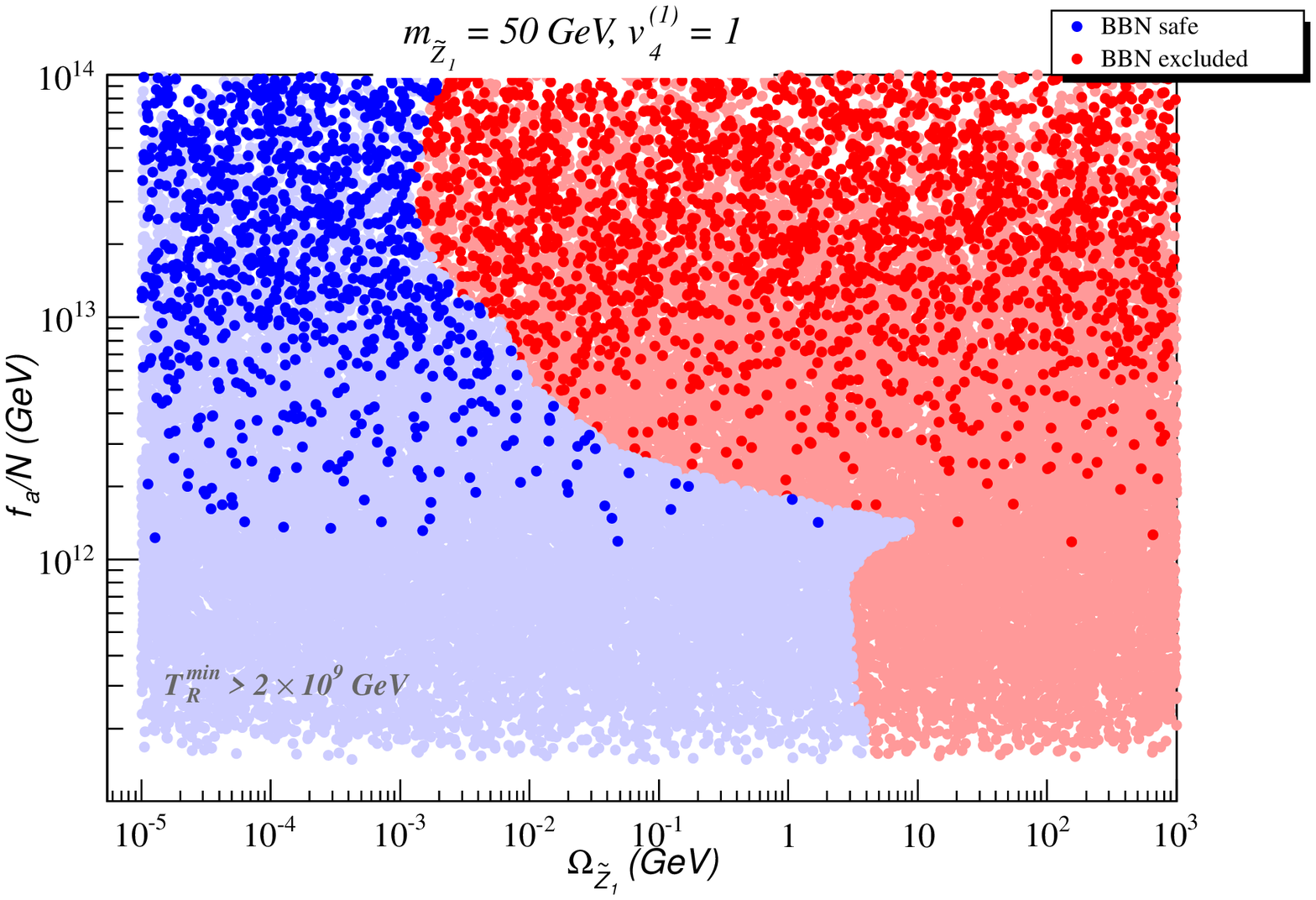} 
\includegraphics[width=10cm]{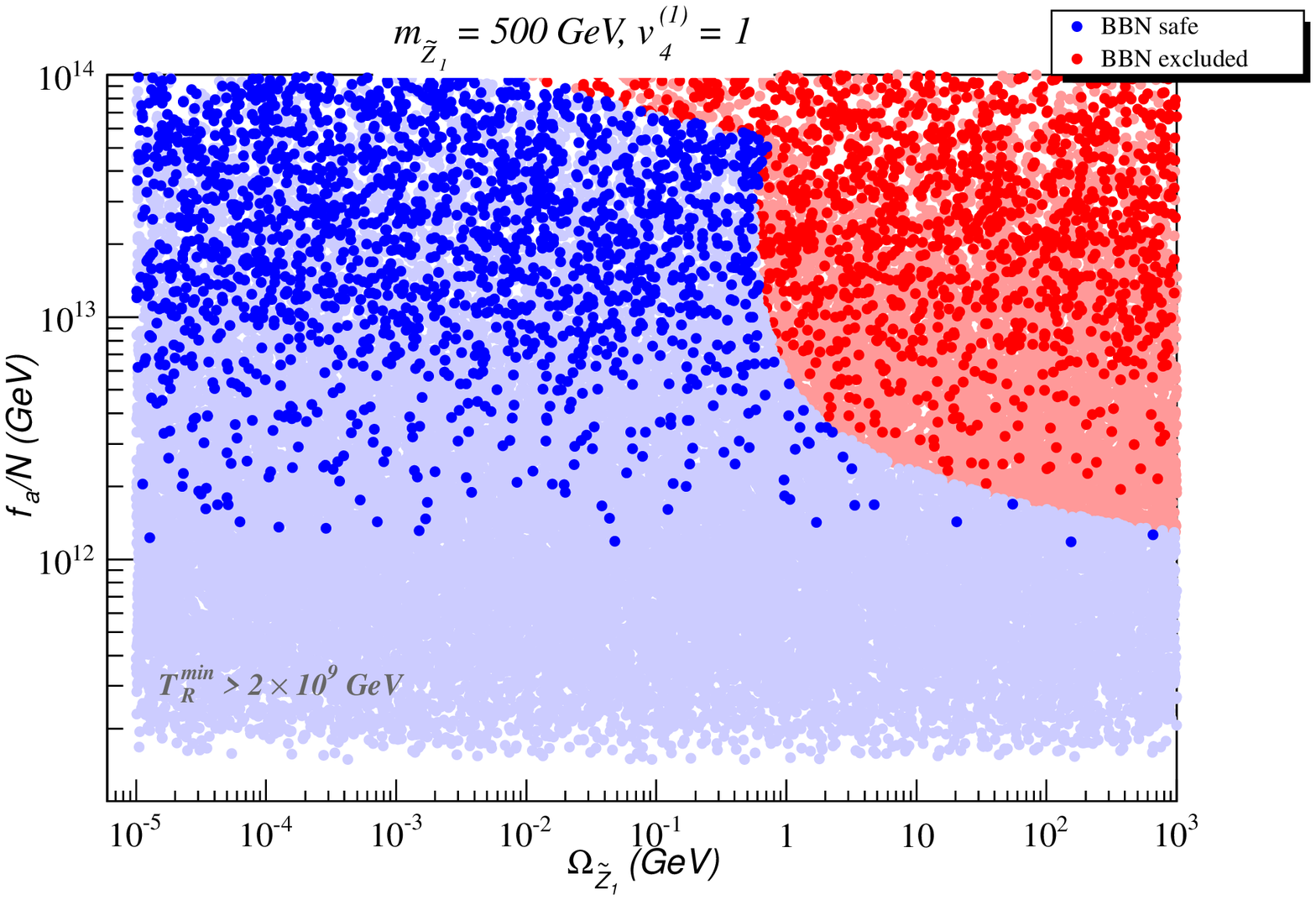}
\caption{Model-independent scatter plot of points with $T_R \gtrsim 2\times10^9$~GeV 
in the $\Omega_{\tz_1} h^2\ vs.\ f_a/N$ plane for $m_{\tz_1}=50$~GeV (upper frame) 
$m_{\tz_1}=500$~GeV (lower frame). Same color code as in Fig.~\ref{fig:BM1fa}.
}\label{fig:Oh2vsFA}}


Experimental consequences of this scenario to reconcile thermal leptogenesis
with the gravitino problem in supersymmetric models include
1.~a discovery at LHC of any of the models discussed here (or others)
which support a gravitino mass in excess of 30 TeV, 2.~the inferred apparent 
relic abundance of neutralinos is typically $\Omega_{\tz_1}h^2\alt 1$, 3.~null 
results from direct or indirect WIMP searches, and 4.~a positive signal for the 
QCD axion at sub-$\mu$eV levels at ADMX~\cite{admx} or other axion detection 
experiments.

\acknowledgments

We thank Laura Covi for helpful discussions. 
We also thank the Galileo Galilei Institute (GGI) for Theoretical Physics 
and the workshop on {\it Dark Matter: Its Origin, Nature and prospects 
for Detection}, for hospitality while this work was initiated.
This research was supported in part by the U.S. Department of Energy,
by the Fulbright Program and CAPES (Brazilian Federal Agency for
Post-Graduate Education), and by the French ANR project {\tt ToolsDMColl}, 
BLAN07-2-194882.

%


\begin{thebibliography}{99}
%
\bibitem{nu_review} For some reviews, see {\it e.g.} V. Barger, D.Marfatia and
K. Whisnant, \ijmpe{12}{2003}{569};
L.Camilleri, E. Lisi and J. Wilkerson, \arnps{58}{2008}{343};
B. Kayser, in C.~Amsler {\it et al.}  [Particle Data Group],
Phys.\ Lett.\  B {\bf 667} (2008) 1.
%
\bibitem{seesaw}  M. Gell-Mann, P. Ramond and R. Slansky, 
in {\it Supergravity, Proceedings of the Workshop}, Stony Brook, NY 1979 
(North-Holland, Amsterdam);  
T. Yanagida, KEK Report No. 79-18, 1979; 
R. Mohapatra and G. Senjanovic,  \prl{44}{1980}{912}.
%
\bibitem{lepto_review}
M. Fukugita and T. Yanagida, \plb{174}{1986}{45};
M. Luty, \prd{45}{1992}{455};
W. Buchm\"uller and M. Plumacher, \plb{389}{1996}{73} and \ijmpa{15}{2000}{5047};
R. Barbieri, P. Creminelli, A. Strumia and N. Tetradis, \npb{575}{2000}{61};
G. F. Giudice, A. Notari, M. Raidal, A. Riotto and A. Strumia, 
\npb{685}{2004}{89};
for a recent review, see W. Buchm\"uller, R. Peccei and T. Yanagida, 
\arnps{55}{2005}{311}.
%
\bibitem{TR} W. Buchm\"uller, P. Di Bari and M. Plumacher, \npb{643}{2002}{367}
and Erratum-ibid,{\bf B793} (2008) 362; \annp{315}{2005}{305} and \njp{6}{2004}{105}.
%
\bibitem{wss} H.~Baer and X.~Tata, {\it Weak Scale Supersymmetry: From 
Superfields to Scattering Events}, 
(Cambridge University Press, 2006).
%
\bibitem{gravprob} S. Weinberg, \prl{48}{1982}{1303}.
%
\bibitem{kl} M. Khlopov and A. Linde, \plb{138}{1984}{265}.  
%
\bibitem{ntlepto} G. Lazarides and Q. Shafi, \plb{258}{1991}{305};
K. Kumekawa, T. Moroi and T. Yanagida, \ptp{92}{1994}{437};
T. Asaka, K. Hamaguchi, M. Kawasaki and T. Yanagida, \plb{464}{1999}{12}.
%
\bibitem{amsb_soln} M. Ibe, R. Kitano, H. Murayama and T. Yanagida, 
\prd{70}{2004}{075012}.
%
\bibitem{covi} W. Buchmuller, L. Covi, K. Hamaguchi, A. Ibarra and T. Yanagida,
\jhep{0703}{2007}{037}
%
\bibitem{covi2}
  L.~Covi, J.~Hasenkamp, S.~Pokorski and J.~Roberts,
\jhep{0911}{2009}{003}.
%
\bibitem{cr_buch} G. Bertone, W. Buchmuller, L. Covi and A. Ibarra, 
JCAP{\bf 0711} (2007) 003; W. Buchmuller, A. Ibarra, T. Shindou, 
F. Takayama and D. Tran, JCAP{\bf 0909} (2009) 021.
%
\bibitem{thooft} G. 't Hooft, \prl{37}{1976}{8}.
%
\bibitem{nedm} For a comprehensive discussion, see 
  M.~Pospelov and A.~Ritz, Annals Phys.\  {\bf 318} (2005) 119.
%
\bibitem{pq} R. Peccei and H. Quinn, \prl{38}{1977}{1440} and \prd{16}{1977}{1791}.
%
\bibitem{ww} S. Weinberg, \prl{40}{1978}{223};
F. Wilczek, \prl{40}{1978}{279}.
%
\bibitem{ksvz} J. E. Kim, \prl{43}{1979}{103};
M. A. Shifman, A. Vainstein and V. I. Zakharov, \npb{166}{1980}{493}.
%
\bibitem{dfsz} M. Dine, W. Fischler and M. Srednicki, \plb{104}{1981}{199};
A. P. Zhitnitskii, \sjp{31}{1980}{260}.
%
\bibitem{bardeen} W. Bardeen and S. H. H. Tye, \plb{74}{1978}{229}. 
%
\bibitem{dicus} D. Dicus, E. Kolb, V. Teplitz and R. Wagoner, \prd{18}{1978}{1829}
and \prd{22}{1980}{839}; for a review, see G. Raffeldt, \hepph{0611350}.
%
\bibitem{sikivie} P. Sikivie, \prl{51}{1983}{1415}.
%
\bibitem{axion} For recent reviews on axion physics, see
M. Turner, \prep{197}{1990}{67};
P. Sikivie, \hepph{0509198};
J. E. Kim and G. Carosi, arXiv:0807.3125.
%
\bibitem{kim} J. E. Kim, \plb{136}{1984}{378}.
%
\bibitem{rtw} K. Rajagopal, M. Turner and F. Wilczek, 
\npb{358}{1991}{447}.
%
\bibitem{amass} E. J. Chun, J. E. Kim and H. P. Nilles, 
\plb{287}{1992}{123}; E. J. Chun and M. Lukas, \plb{357}{1995}{43}.
%
\bibitem{ckkr} L. Covi, J. E. Kim and L. Roszkowski, \prl{82}{1999}{4180}; 
L. Covi, H. B. Kim, J. E. Kim and L. Roszkowski, \jhep{0105}{2001}{033}.
%
\bibitem{axino} For recent reviews of axino dark matter, see
F. Steffen, \epjc{59}{2009}{557}; L. Covi and J. E. Kim, 
\njp{11}{2009}{105003}.
%
\bibitem{st} R. J. Scherrer and M. Turner, \prd{31}{1985}{681};
G. Lazarides, R. Schaefer, D. Seckel and Q. Shafi, \npb{346}{1990}{193};
J. E. Kim, \prl{67}{1991}{3465}.
%
\bibitem{hasenkamp}
  J.~Hasenkamp and J.~Kersten, arXiv:1008.1740 [hep-ph].
%
\bibitem{ay} T. Asaka and T. Yanagida, \plb{494}{2000}{297}.
%
\bibitem{nilles} H. P. Nilles and S. Raby, \npb{198}{1982}{102}.
%
\bibitem{wmap7} E. Komatsu {\it et al.} (WMAP collaboration), 
arXiv:1001.4538 (2010).
%
\bibitem{vacmis} L. F. Abbott and P. Sikivie, \plb{120}{1983}{133};
J. Preskill, M. Wise and F. Wilczek, \plb{120}{1983}{127};
M. Dine and W. Fischler, \plb{120}{1983}{137};
M. Turner, \prd{33}{1986}{889}; L. Visinelli and P. Gondolo, 
\prd{80}{2009}{035024}.
%
\bibitem{steffen} A. Brandenburg and F. Steffen, JCAP{\bf 0408} (2004) 008.
%
\bibitem{strumia} A. Strumia, \jhep{1006}{2010}{036}.
%
\bibitem{isared} H. Baer, C. Balazs and A.Belyaev, \jhep{0203}{2002}{042}.
%
\bibitem{isatools} H. Baer, C. Balazs, A.Belyaev, J. K. Mizukoshi, 
X. Tata and Y. Wang, \jhep{0207}{2002}{050}.
%
\bibitem{isajet} F. Paige, S. Protopopescu, H. Baer and X. Tata, \hepph{0312045}; 
http://www.nhn.ou.edu/$\sim$isajet/
%
\bibitem{relic_G} M. Bolz, A. Brandenburg and W. Buchmuller, \npb{606}{2001}{518};
J. Pradler and F. Steffen, \prd{75}{2007}{023509};
V. S. Rychkov and A. Strumia, \prd{75}{2007}{075011}.
%
\bibitem{axdm} H. Baer, A. Box and H. Summy, 
\jhep{0908}{2009}{080}.
%
\bibitem{ellis} 
R. H. Cyburt, J. Ellis, B. D. Fields and K. A. Olive, \prd{67}{2003}{103521};
R. H. Cyburt, J. Ellis, B. D. Fields, F. Luo, K. Olive and V. Spanos, 
JCAP{\bf 0910} (2009) 021.
%
\bibitem{kohri} M. Kawasaki, K. Kohri and T. Moroi, 
\plb{625}{2005}{7} and \prd{71}{2005}{083502};
K. Kohri, T. Moroi and A. Yotsuyanagi, \prd{73}{2006}{123511};
for an update, see M. Kawasaki, K. Kohri, T. Moroi and A. Yotsuyanagi, 
arXiv:0804.3745 (2008).
%
\bibitem{jedamzik} K. Jedamzik, \prd{70}{2004}{063524}
and \prd{74}{2006}{103509}.
%
\bibitem{ax19} H. Baer and A. Box, \epjc{68}{2010}{523};
H. Baer, A. Box and H. Summy, arXiv:1005.2215 (2010).
%
\bibitem{hb} H. Baer and J. Ferrandis, \prl{87}{2001}{211803};
D. Auto, H. Baer, C. Balazs, A. Belyaev, J. Ferrandis 
and X. Tata, \jhep{0306}{2003}{023}.
%
\bibitem{bdr} T. Blazek, R. Dermisek and S. Raby, \prl{88}{2002}{111804} and
\prd{65}{2002}{115004}; R. Dermisek, S. Raby, L. Roszkowski and
R. Ruiz de Austri, \jhep{0304}{2003}{037} and \jhep{0509}{2005}{029};
M. Albrecht, W. Altmannshofer, A. Buras, D. Guadagnoli and D. Straub,
\jhep{0710}{2007}{055}; W. Altmannshofer, D. Guadagnoli, S. Raby and
D. Straub, arXiv:0801.4363 (2008); D. Guadagnoli, arXiv:0810.0450 (2008).
%
\bibitem{bkss} H. Baer, S. Kraml, S. Sekmen and H. Summy, \jhep{0803}{2008}{056}.
%
\bibitem{imh} J. Feng, C. Kolda and N. Polonsky, \npb{546}{1999}{3}; 
J. Bagger, J. Feng and N. Polonsky, \npb{563}{1999}{3};
J. Bagger, J. Feng, N. Polonsky and R. Zhang, \plb{473}{2000}{264};
H. Baer,P. Mercadante and X. Tata, \plb{475}{2000}{289};
H. Baer, C. Balazs, M. Brhlik, P. Mercadante, X. Tata and Y. Wang,
\prd{64}{2001}{015002}; see also H. Baer, M. Diaz, P. Quintana and X. Tata, 
\jhep{0004}{2000}{016}.
%
\bibitem{bs} H. Baer and H. Summy, \plb{666}{2008}{5}.
%
\bibitem{bhkss} H. Baer, M. Haider, S. Kraml, S. Sekmen and H. Summy, 
JCAP{\bf 0902} (2009) 002.
%
\bibitem{auto} D. Auto, H. Baer, A. Belyaev and T. Krupovnickas, 
\jhep{0410}{2004}{066}.
%
\bibitem{lhc7} H. Baer, S. Kraml, A. Lessa and S. Sekmen, 
\jhep{1002}{2010}{055}.
%
\bibitem{ckn} A. Cohen, D. Kaplan and A. Nelson, 
\plb{388}{1996}{588}.
%
\bibitem{esusy} H. Baer, S. Kraml, A. Lessa, S. Sekmen and X. Tata,
arXiv:1007.3897 (2010).
%
\bibitem{amsb} L. Randall and R. Sundrum, \npb{557}{1999}{79};
G. Giudice, M. Luty, H. Murayama and R. Rattazzi, \jhep{9812}{1998}{027}.
%
\bibitem{shanta} S. P. de Alwis, \jhep{1003}{2010}{078};
H. Baer, S. P. de Alwis, K. Givens, S. Rajagopalan and H. Summy,
\jhep{1005}{2010}{069}.
%
\bibitem{hcamsb} R. Dermisek, H. Verlinde and L. T. Wang, 
\prl{100}{2008}{131804}; H. Baer, R. Dermisek, S. Rajagopalan and H. Summy, 
\jhep{0910}{2009}{078}.
%
\bibitem{shibi} H. Baer, R. Dermisek, S. Rajagopalan and H. Summy, 
JCAP{\bf 1007} (2010) 014.
%
\bibitem{mmamsb} K. Choi, A. Falkowski, H. P. Nilles, M. Olechowski and
S. Pokorski, \jhep{0411}{2004}{076}; K. Choi, A. Falkowski, H. P. Nilles
and M. Olechowski, \npb{718}{2005}{113}; K. Choi, K-S. Jeong,
  \jhep{0701}{2007}{103}; A. Falkowski, O. Lebedev and Y. Mambrini,
  \jhep{0511}{2005}{034}; H. Baer, E. Park, X. Tata and T. T. Wang, 
\jhep{0608}{2006}{041}; \plb{641}{2006}{447}; \jhep{0706}{2007}{033}.
%
\bibitem{kklt} S. Kachru, R. Kallosh, A. Linde and S. P. Trivedi,
\prd{68}{2003}{046005}. 
%
\bibitem{bwca} H. Baer, T. Krupovnickas, A. Mustafayev, E. K. Park, S. Profumo and X. Tata,
\jhep{0512}{2005}{011}.
%
\bibitem{Afunnel} M.~Drees and M.~Nojiri, \prd{47}{1993}{376}; 
H.~Baer and M.~Brhlik, \prd{57}{1998}{567};
H.~Baer, M.~Brhlik, M.~Diaz, J.~Ferrandis, P.~Mercadante,
P.~Quintana and X.~Tata, \prd{63}{2001}{015007};
J.~Ellis, T.~Falk, G.~Ganis, K.~Olive and M.~Srednicki, \plb{510}{2001}{236}; 
L.~Roszkowski, R.~Ruiz de Austri and T.~Nihei, \jhep{0108}{2001}{024}; 
A.~Djouadi, M.~Drees and J.~L.~Kneur, \jhep{0108}{2001}{055}; 
A.~Lahanas and V.~Spanos, \epjc{23}{2002}{185}.
%
\bibitem{jlm} K. Jedamzik, M. LeMoine and G. Moultaka,
JCAP{\bf 0607} (2006) 010.
%
\bibitem{warm}
  M.~Viel, G.~D.~Becker, J.~S.~Bolton, M.~G.~Haehnelt, M.~Rauch and W.~L.~W.~Sargent,
  Phys.\ Rev.\ Lett.\  {\bf 100} (2008) 041304;
  A.~V.~Maccio' and F.~Fontanot,  arXiv:0910.2460 [astro-ph.CO]; 
  A.~Boyarsky, J.~Lesgourgues, O.~Ruchayskiy and M.~Viel, JCAP {\bf 0905} (2009) 012; 
  D.~Boyanovsky and J.~Wu, arXiv:1008.0992 [astro-ph.CO].
%
\bibitem{hot}
  S.~Hannestad, A.~Mirizzi, G.~G.~Raffelt and Y.~Y.~Y.~Wong,
  JCAP {\bf 1008} (2010) 001.
%
\bibitem{admx} 
L. Duffy {\it et al.}, \prl{95}{2005}{091304} and \prd{74}{2006}{012006};
for a review, see S. Asztalos, L. Rosenberg, K. van Bibber, P. Sikivie
and K. Zioutas, \arnps{56}{2006}{293}.
%
\bibitem{dine} M. Dine, A. Kagan and S. Samuel, \plb{243}{1990}{250}.
%
\end{thebibliography}
\end{document}